\documentclass[manuscript,screen]{acmart}

\AtBeginDocument{%
  }

\setcopyright{acmlicensed}
\copyrightyear{2025}
\acmYear{2025}
\acmDOI{XXXXXXX.XXXXXXX}

\acmConference[CSCW'26]{Proc. ACM Hum.-Comput. Interact. CSCW}

\usepackage{float}

\usepackage{tikz}
\usetikzlibrary{calc, shapes.arrows, positioning, backgrounds, patterns, arrows.meta}

\definecolor{headerGray}{HTML}{737F89}
\definecolor{bgRow1}{HTML}{DEEAF3} 
\definecolor{bgRow2}{HTML}{C5DDF3} 
\definecolor{bgRow3}{HTML}{90C2F2} 
\definecolor{bgRow4}{HTML}{5FA8EF} 
\definecolor{arrowWhite}{HTML}{FFFFFF}
\definecolor{textDark}{HTML}{222222}

\begin{document}

\title{AI Didn't Start the Fire: Examining the Stack Exchange Moderator and Contributor Strike}


\author{Yiwei Wu}
 \email{yiweiwu@austin.utexas.edu}
 \orcid{0009-0005-0871-9225}
 \affiliation{
  \institution{School of Information, The University of Texas at Austin}
  \city{Austin}
   \state{TX}
 \country{USA}
 }

 \author{Leah Ajmani}
 \email{ajman004@umn.edu}
 \orcid{0000-0003-2468-6070}
 \affiliation{
  \institution{University of Minnesota}
  \city{Minneapolis}
   \state{MN}
 \country{USA}
 }

\author{Nathan TeBlunthuis}
\email{nathante@utexas.edu}
\orcid{0000-0002-3333-5013}
 \affiliation{
  \institution{School of Information, The University of Texas at Austin}
  \city{Austin}
   \state{TX}
 \country{USA}
 }

 \author{Hanlin Li}
 \email{lihanlin@utexas.edu}
 \orcid{0000-0002-0688-9918}
 \affiliation{
  \institution{School of Information, The University of Texas at Austin}
  \city{Austin}
   \state{TX}
 \country{USA}
 }
\renewcommand{\shortauthors}{Wu et al.}

\begin{abstract}
Online communities and their host platforms are mutually dependent yet conflict-prone. When platform policies clash with community values, communities have resisted through strikes, blackouts, and even migration to other platforms. Through such collective actions, communities have sometimes won concessions but these have frequently proved temporary.  Prior research has investigated strike events and migration chains, but the processes by which community-platform conflict unfolds remain obscure. How do community-platform relationships deteriorate? How do communities organize collective action? How do participants proceed in the aftermath?
We investigate a conflict between the Stack Exchange platform and community that occurred in 2023 around an emergency arising from the release of large language models (LLMs). Based on a qualitative thematic analysis of 2,070 messages on Meta Stack Exchange and 14 interviews with community members, we surface how the 2023 conflict was preceded by a long-term deterioration in the community-platform relationship driven in particular by the platform's disregard for the community's highly-valued participatory role in governance. Moreover, the platform's policy response to LLMs aggravated the community's sense of crisis triggering the strike mobilization.  We analyze how the mobilization was coordinated through a tiered leadership and communication structure, as well as how community members pivoted in the aftermath. Building on recent theoretical scholarship in social computing, we use Hirshman's exit, voice and loyalty framework to theorize the challenges of community-platform relations evinced in our data. Finally, we recommend ways that platforms and communities can institute participatory governance to be durable and effective.



\end{abstract}



\received{May 2025}

\begin{teaserfigure}
\centering
\sffamily
\def\canvasWidth{22}
\def\colOSplit{6.3}
\def\colSplit{13}     
\def\colCenter{17.5}    

\def\yRowOne{-5.1}
\def\yRowTwo{-8.5}
\def\yRowThree{-11.9}
\def\yRowFour{-15.4}
\def\parMiniPageWidth{4.7cm}
\def\arrowDepth{1.2} 
\def\smallArrowDepth{0.6} 
\def\arrowStart{\colCenter - \arrowDepth}   
\def\arrowEnd{\colCenter + \arrowDepth}   
\def\smallArrowStart{\colCenter - \smallArrowDepth}   
\def\smallArrowEnd{\colCenter + \smallArrowDepth}   
\def\rowThreeArrowDepth{0.5}

\def\tinyArrowInnerWidth{0.25}
\def\tinyArrowOffset{2}
\def\tinyArrowVertOffset{0.3}
\def\tinyArrowTwoCenter{\colCenter + \tinyArrowOffset}
\def\tinyArrowOneCenter{\colCenter - \tinyArrowOffset}
\def\tinyArrowWidth{0.5}
\def\tinyArrowDepth{\tinyArrowWidth}
\def\tinyArrowOneStart{\tinyArrowOneCenter + \tinyArrowWidth}   
\def\tinyArrowTwoStart{\tinyArrowTwoCenter + \tinyArrowWidth}   
\def\tinyArrowOneEnd{\tinyArrowOneCenter - \tinyArrowWidth}   
\def\tinyArrowTwoEnd{\tinyArrowTwoCenter - \tinyArrowWidth}   

\begin{tikzpicture}[
    scale=0.68,
    node distance=0cm,
    outer sep=0pt,
    headerText/.style={white, font=\bfseries, align=center},
    colTitle/.style={textDark, font=\bfseries\small},
    bulletItem/.style={textDark, font=\footnotesize, align=left, anchor=north west, text width=5.2cm, xshift=0.3cm},
    parItem/.style={textDark, font=\footnotesize, align=left, anchor=north west, text width=\parMiniPageWidth},
    dateItem/.style={textDark, font=\footnotesize, align=left, anchor=north west, text width=3.7cm},
    theoryTitle/.style={textDark, font=\bfseries\Large, anchor=north, align=center, text width=4cm},
    theoryText/.style={textDark, font=\footnotesize\bfseries, align=center, text width=4.5cm},
    fineDashed/.style={draw=bgRow3, line width=1.1pt},
    theoryArrow/.style={draw=bgRow1, fill opacity=0, line width=1pt},
    theoryArrow2/.style={draw=bgRow1,fill=bgRow1, fill opacity=0, line width=9},
    ]


\fill[bgRow4] (0,0) rectangle (\canvasWidth, \yRowFour);

\fill[bgRow2] 
    (0,0) -- 
    (0, \yRowTwo) -- 
    (\canvasWidth, \yRowTwo) -- 
    (\canvasWidth, 0) -- cycle;
    
\fill[bgRow3]
    (0,\yRowTwo) -- 
    (0, \yRowThree) -- 
    (\tinyArrowOneStart -\tinyArrowInnerWidth, \yRowThree) -- 
    (\tinyArrowOneStart - \tinyArrowInnerWidth, \yRowThree - \tinyArrowVertOffset) --
    (\tinyArrowOneStart, \yRowThree - \tinyArrowVertOffset) --
    (\tinyArrowOneCenter, \yRowThree - \tinyArrowVertOffset - \tinyArrowDepth) --  
    (\tinyArrowOneEnd, \yRowThree - \tinyArrowVertOffset) --
    (\tinyArrowOneEnd + \tinyArrowInnerWidth, \yRowThree - \tinyArrowVertOffset) --
    (\tinyArrowOneEnd + \tinyArrowInnerWidth, \yRowThree) --
    (\tinyArrowTwoStart -\tinyArrowInnerWidth, \yRowThree) -- 
    (\tinyArrowTwoStart - \tinyArrowInnerWidth, \yRowThree - \tinyArrowVertOffset) --
    (\tinyArrowTwoStart, \yRowThree - \tinyArrowVertOffset) --
    (\tinyArrowTwoCenter, \yRowThree - \tinyArrowVertOffset - \tinyArrowDepth) -- 
    (\tinyArrowTwoEnd, \yRowThree - \tinyArrowVertOffset) --
    (\tinyArrowTwoEnd + \tinyArrowInnerWidth, \yRowThree - \tinyArrowVertOffset) --
    (\tinyArrowTwoEnd + \tinyArrowInnerWidth, \yRowThree) --
    (\canvasWidth, \yRowThree) -- 
    (\canvasWidth, \yRowTwo) -- cycle;


\fill[bgRow2]
    (\tinyArrowOneStart - 0.5 * \tinyArrowInnerWidth, \yRowTwo+0.1) -- 
    (\tinyArrowOneCenter, \yRowTwo - \tinyArrowInnerWidth) -- 
    (\tinyArrowOneEnd + 0.5 * \tinyArrowInnerWidth, \yRowTwo+0.1) -- 
    (\smallArrowStart, \yRowTwo+0.1) --
    (\smallArrowStart, \yRowTwo - \rowThreeArrowDepth) --
    (\arrowStart, \yRowTwo - \rowThreeArrowDepth) -- 
    (\colCenter, \yRowTwo - \arrowDepth - \rowThreeArrowDepth) -- 
    (\arrowEnd, \yRowTwo - \rowThreeArrowDepth) --
    (\smallArrowEnd, \yRowTwo - \rowThreeArrowDepth) --
    (\smallArrowEnd, \yRowTwo+0.1) -- 
    (\tinyArrowTwoStart - 0.5 * \tinyArrowInnerWidth, \yRowTwo+0.1) -- 
    (\tinyArrowTwoCenter, \yRowTwo - \tinyArrowInnerWidth) -- 
    (\tinyArrowTwoEnd + 0.5 * \tinyArrowInnerWidth, \yRowTwo+0.1) -- cycle;


\fill[bgRow1] 
    (0,0) -- 
    (0, \yRowOne) -- 
    (\smallArrowStart, \yRowOne) -- 
    (\colCenter, \yRowOne - \smallArrowDepth) -- 
    (\smallArrowEnd, \yRowOne) -- 
    (\canvasWidth, \yRowOne) -- 
    (\canvasWidth, 0) -- cycle;

\fill[headerGray] (0,0) rectangle (\canvasWidth, 1.2);


\node[headerText] at (2.5, 0.6) {Timeline};
\node[headerText] at (8.5, 0.6) {Themes};
\node[headerText] at (\colCenter, 0.6) {Theoretical Interpretation};

\node[dateItem] at (0.5, -0.25) {\textbf{2009:} SE founded with a community-first moderation model};
\node[dateItem] at (0.5, -2) {\textbf{2019:} ``Monica incident'' erodes trust in company};

\node[parItem] at (\colOSplit, -0.25){\begin{minipage}{\parMiniPageWidth}
    \hspace{1.5em}\textbf{Longstanding Grievances}
    \begin{itemize}
    \item Governance shifted from participatory to unilateral
    \item SE, Inc. lacks clarity and transparency in communication
    \item A history of grievances set the stage for conflict
    \item The SE community demands participatory governance
    \end{itemize}
  \end{minipage}};


\node[dateItem] at (0.5, \yRowOne-0.2) {\textbf{5/29/23:} SE, Inc. privately asks mods to stop flagging AI content};
\node[dateItem] at (0.5, \yRowOne-1.25) {\textbf{5/30/23:} Public AI policy conflicts with version sent to mods};

\node[parItem] at (\colOSplit, \yRowOne - 0.2) {\begin{minipage}{\parMiniPageWidth}
    \hspace{1.5em} \textbf{Emergent Grievances}
    \begin{itemize}
    \item Low-quality AI-generated content floods the site
    \item Shutting down the data dump violated the free knowledge ethos
    \item AI poses attribution issues
    \end{itemize}
    \end{minipage}
};

\node[dateItem] at (0.5, \yRowTwo-0.2) {\textbf{6/1/23:} Mods coordinated strike via private Discord};
\node[dateItem] at (0.5, \yRowTwo-1.25) {\textbf{6/5/23:} Community petition launched};
\node[dateItem] at (0.5, \yRowTwo-2.3) {\textbf{6/7/23:} SE CEO pushed to shut down data dump};

\node[parItem] at (\colOSplit, \yRowTwo-0.2) {\begin{minipage}{\parMiniPageWidth}
    \hspace{1.5em}\textbf{Building Collective Power}
    \begin{itemize}
    \item The SE community organized tiered collective action
    \item The SE community withheld contributions and moderation
    \end{itemize}
  \end{minipage}};

\node[colTitle, anchor=north, align=center, text width=1.7cm] (voiceTitle) at (\colCenter-2,\yRowTwo + 2.5) {Expressing Voice};

\node[colTitle, align=center, anchor=north, text width=1.7cm] (exitTitle) at (\colCenter+2,\yRowTwo + 2.5) {Exiting the Platform};

\node[theoryText, text width = 2cm](pursuingEffective) at (\colCenter, \yRowTwo - \arrowDepth - \rowThreeArrowDepth - 0.5) {Pursuing Effective Voice};

\node[parItem] (bindingMechanisms) at (\colCenter-3.4,\yRowThree-0.8){\begin{minipage}{2.5cm}
    \textbf{Binding Mechanisms:}

     Formalize commitments to community input
     
  \end{minipage}};

\node[parItem] (credibleExit) at (\colCenter+0.4, \yRowThree-0.8) {\begin{minipage}{2.4cm}

    \textbf{Credible Exit:}

    Provide contributors with functional alternatives
    \end{minipage}
    };

\node[dateItem] at (0.5, \yRowThree-0.2) {\textbf{8/7/2023 Onwards}: Migration to Codidact/TopAnswers continues.  Community members archive SE data};

\node[parItem] at (\colOSplit, \yRowThree-0.2) {\begin{minipage}{\parMiniPageWidth}
    \hspace{1.5em} \textbf{Migration and Preservation}
    \begin{itemize}
    \item The SE community seeks alternatives through migration
    \item The SE community shares decentralized knowledge through archival activism
    \end{itemize}
    \end{minipage}};

\node[] (arrowLeftEnd) at (\colCenter-0.5,\yRowTwo - 1) {}; 
\node[] (arrowRightEnd) at (\colCenter+0.5,\yRowTwo - 1) {}; 
\node[] (arrowRightTopEnd) at (\colCenter+2,\yRowOne+0.5) {}; 


\draw[bgRow4, line width=1.1pt]
        (\tinyArrowOneStart - \tinyArrowInnerWidth, \yRowThree - \tinyArrowVertOffset)--
       (\tinyArrowOneStart - \tinyArrowInnerWidth, \yRowTwo + 1.2) --
        (\tinyArrowOneEnd + \tinyArrowInnerWidth, \yRowTwo + 1.2) --
        (\tinyArrowOneEnd + \tinyArrowInnerWidth, \yRowThree - \tinyArrowVertOffset)
                ;

\draw[bgRow4, line width=1.1pt]
        (\tinyArrowTwoStart - \tinyArrowInnerWidth, \yRowThree - \tinyArrowVertOffset)--
       (\tinyArrowTwoStart - \tinyArrowInnerWidth, \yRowTwo + 1.2) --
        (\tinyArrowTwoEnd + \tinyArrowInnerWidth, \yRowTwo + 1.2) --
        (\tinyArrowTwoEnd + \tinyArrowInnerWidth, \yRowThree - \tinyArrowVertOffset)
                ;


\draw[fineDashed]
    (\smallArrowEnd, \yRowOne+2) --   
    (\smallArrowStart, \yRowOne+2) --   
    (\smallArrowStart, \yRowTwo-\rowThreeArrowDepth); 

\draw[fineDashed] 
    (\smallArrowStart, \yRowOne+2) --   
    (\smallArrowEnd, \yRowOne+2) --   
    (\smallArrowEnd, \yRowTwo-\rowThreeArrowDepth); --   

\node[theoryTitle,font=\huge\bfseries] at (\colCenter, -0.8) {Declining Loyalty};

\end{tikzpicture}

\vspace{-2em}
\caption{How historical tensions and strike-related events on Stack Exchange align with the SE community’s grievances, their actions, and our theoretical interpretations of loyalty, voice, and exit.}
\label{fig:timeline}
\end{teaserfigure}
\maketitle

\section{Introduction}

Due to the impact of generative AI technologies on online communities, numerous conflicts have erupted between content platforms such as Reddit, Stack Exchange, Tumblr, and Deviantart and the online communities who produce and share valuable knowledge and cultural works through them.\footnote{For documentation of conflicts see \url{https://slate.com/technology/2024/05/deviantart-what-happened-ai-decline-lawsuit-stability.html} for Deviantart, \url{https://www.tumblr.com/staff/743510217982083072/hi-tumblr-its-tumblr-were-working-on-some} for Tumblr, and \url{https://en.wikipedia.org/wiki/2023_Reddit_API_controversy} for Reddit.} 
As platforms sell user-generated content to AI developers, contributors are concerned that their communities will be harmed both by such use and by a deluge of AI-generated content. We investigate a prominent case in which a community advocated to protect themselves from such threats and to oppose their platforms' unilateral policy changes related to AI: the 2023 Stack Exchange moderator and contributor strike.

In June 2023, Stack Exchange (SE), an online knowledge sharing platform, experienced a strike by contributors and moderators against policy  and operational changes imposed by the platform company (SE, Inc.), including new policies around moderation of AI-generated content and restrictions on API and data dump access~\citep{mithical_moderation_2023}. SE, known for over 15 years of collaboratively curated technical Q\&A content, has a relatively participatory community governance structure, including election for moderators, an open process of earning moderation privileges, and reputation-based authority. Nevertheless, relations between SE, Inc. and many of the most dedicated SE contributors broke down, disrupting ongoing collaboration and arguably threatening the platform's sustainability.

Relationships between online communities and their host platforms are often contentious, even before AI technologies posed new threats. 
How can these relationships be less acrimonious? Studies in social computing have investigated how online communities take collective action, such as protest or migration, as a recourse when platforms violate their expectations and values~\citep[e.g.,][]{matias_going_2016, centivany_popcorn_2016, fiesler_moving_2020, waltenberger_digital_2025}. Such investigations tend to focus on collective action \emph{during a conflict}, such as a protest against a platform policy change. However, during a conflict is a difficult time to implement structures, designs, or mechanisms that the community or platform need to resolve their differences.  For instance, ~\citet{frey_effective_2023} draw from Albert O. Hirshman's influential exit, voice, and loyalty framework~\citep{hirschman_exit_1970}, to theorize that \textit{effective voice} mechanisms might (legally, technically, or otherwise) bind a platform to give the community a role in decision-making. Assured that the platform must account for their needs, such a community may have less need to strike or migrate in a conflict. Moreover, in Hirshman's framework, collective action (i.e., a strike or migration) comes as a last resort only after expressing voice (i.e., communication with the platform) has proved ineffective and their loyalty (i.e., commitment to the platform) has been lost. To better understand the potential of proposals such as effective voice, we must in turn understand the long-term processes by which a once-harmonious relationship breaks down. We thus asked the following research questions about the SE strike:


\textbf{RQ1}: What long-standing and emergent grievances from the community led to the strike? 

\textbf{RQ2}: How was the strike organized?

\textbf{RQ3}: What happened in the aftermath of the strike?

We collected data from mixed sources to gain a comprehensive understanding of the strike event. We first gathered online discussions (both posts and comments) from Meta Stack Exchange (the site on SE designated for policy discussions) and used these to design our interview questions. We then conducted 14 semi-structured interviews with current and former Stack Exchange users, moderators, and employees supplemented with close readings of online discussions around the strike to understand why and how the strike took place. 
Our analysis reveals how despite SE's affordances for participatory governance, SE, Inc. lost community loyalty by eroding the community's governance role, acting unilaterally, and communicating opaquely in the years before the strike. Generative AI indeed sparked an emergency for the community, but this context of lost loyalty was inextricable from the community's reasons to strike. In the course of the strike, the communities organized contributors and moderators using a tiered structure to coordinate the withdrawal of volunteer moderation and contributions.  After the strike, we found that contributors sought alternatives and engaged in archival activism to safeguard SE's knowledge. We thus provide novel qualitative evidence about how online communities experience exit, voice, and loyalty dynamics as they navigate tensions with the platform over time and in emergencies. We summarized a timeline of events along with our findings and theoretical interpretations in Figure \ref{fig:timeline}.

Our study makes several contributions to the CSCW and social computing literature on online communities and participatory governance. First, we conduct a novel investigation of a major episode of community-platform contention linked to the emergence of generative AI technologies. Our results contribute to the emerging literature in social computing on participatory governance~\citep{jhaver_decentralizing_2023, zuckerman_community_2023, frey_effective_2023, schneider_admins_2022, frey_this_2019, kuo_policycraft_2025} and highlight communities' desire to govern the output of their data labor that creates and maintains valuable knowledge and information~\citep{li_dimensions_2023,vincent_data_2021}.Second, to our knowledge, we present the first qualitative evidence of exit, voice, and loyalty dynamics as faced by contributors to knowledge projects. In doing so, we show that even for platforms with extensive features to support community-driven moderation, governance crisis can still occur. Finally, we make concrete design and policy recommendations to manage such crises: Online communities, platforms, and AI companies ought to institute participatory governance by supporting effective voice through binding mechanisms and credible exit to functional alternatives.

\section{Background and Related Work}

Below, we introduce the lines of inquiry that inspired and motivated our study and the orienting concepts that framed our qualitative analysis~\citep{charmaz_constructing_2006}. First, we focus on the specific actions of content moderation in community governance and discuss relevant literature on labor of care in online communities. We then discuss participatory governance---a bottom-up process for decision making, contrary to content moderation---and position our study amid the growing CSCW literature on this topic.  Finally, we review recent work on the problems generative-AI technologies pose to online communities.

\subsection{Moderation and Online Community Governance}\label{background:moderation}

Large online communities with complex projects, such as SE, necessitate complex governance procedures and roles. Following ~\citet{grimmelmann_virtues_2015}, governance is the \textit{``mechanisms that structure participation in a community to facilitate cooperation and prevent abuse.''} Thus, governance is not just removing harmful content, but about fostering safety and cooperation. Similarly, governance and moderation are often studied in CSCW and HCI through the lens of community health~\citep{seering_metaphors_2022}. For example,~\citet{seering_metaphors_2022} note that moderators' roles are not just custodial, but often nurturing akin to curation or gardening.~\citet{Gilbert2020-hu} describes such roles as essential to building the r/AskHistorians community. In this vein, recent CSCW work has reconceptualized moderation from preventing harms to caring for a community~\cite{Gilbert2023-pg, mcinnisReportingCommunityBeat2021}. 

In addition, a growing literature foregrounds the \emph{labor} of community governance. Particularly true in peer-production contexts, such as Wikipedia and Stack Exchange where a community collaborates on an artifact, moderators expend considerable valuable effort~\citep{wang_quality_2015,dabbish_transparency_2014, vincent_measuring_2019, li_all_2022}. For example, banning vandals from Wikipedia can involve complex coordinated efforts as well as complex tools~\cite{Geiger2010-ny, teblunthuis_effects_2021}. Moderators often find this labor taxing, especially in volunteer contexts~\cite{Gilbert2020-hu}. 
Moreover, as moderator labor is often extracted in for-profit ways, often without moderators' awareness,
it can be seen as invisible work~\cite{li_all_2022}. For example,~\citet{liMeasuringMonetaryValue2022} found that Reddit moderator labor was worth a minimum of 3.4 million USD in 2020 based on the median hourly wage for comparable services. Similarly,~\citet{Vincent2021-xl} found that Wikipedians' labor often benefits search engines without any form of compensation. As we discuss below in §\ref{background:ai}, the recent rise of large language models (LLMs) poses additional burdens to SE's overworked, unpaid moderators. 
 
\subsection{Participatory Governance and Effective Voice}

While the literature on online community moderation and governance discussed in §\ref{background:moderation} usefully foregrounds the care work of moderation, it nevertheless tends to view governance and moderation as a ``top-down'' process. Indeed, many online communities, even in peer production, can be characterized by ``implicit feudalism", where decision-making power is concentrated in a ``benevolent dictator for life'', or a small group of moderators ~\cite{schneider_admins_2022, shaw_laboratories_2014, frey_effective_2023}.
By contrast, the participatory governance perspective calls for distributing governance power among contributors through bottom-up mechanisms such as distributed moderation, elections for leadership roles, open policy discussions, and feedback systems ~\cite{jhaver_decentralizing_2023, schneider_governable_2024, matias_civilservant_2018, zhang_policykit_2020, weld_making_2024, gowder_networked_2023}.
In this vein, HCI and social computing scholars have  developed systems for participatory governance, such as CivilServant and PolicyCraft ~\cite{matias_civilservant_2018,kuo_policycraft_2024,zhang_policykit_2020}. Such tools support communities in enacting governance policies that devolve and decentralize administrative control. In the broadest sense, participatory governance encompasses all forms of decision making within a community, not limited to content moderation, community norms, and policies.
As we unpack in §\ref{sec:findings} SE community members intensely pursued increases in participatory decision making. They did so even though, as we discuss in §\ref{empiricalSetting}, governance on SE was already comparatively participatory.

Our study addresses two interrelated challenges for participatory governance: (1) platforms' corporate strategies have little role for  meaningful community engagement, and (2) contributors’ limited influence over platform decisions~\citep{frey_effective_2023, gowder_networked_2023}. 
Highlighting such challenges, ~\citet{gowder_networked_2023}, argues that social media platforms often lack institutional mechanisms both to understand community needs and to commit to beneficent governance policies in the long-term. Similarly, 
~\citet{spaa_understanding_2019} emphasize how community input fails to translate into platform policy due to misalignment between community knowledge and institutional norms ~\cite{spaa_understanding_2019}. 
Without formalized structures for bottom-up change, participatory governance risks being symbolic rather than substantive.
Such problems may become especially salient during governance crises, when sudden unilateral decisions are cast on the community ~\cite{centivany_popcorn_2016}. Thus, fully realized participatory governance requires not just surface-level mechanisms (e.g., feedback forms, user polls), but also formal inclusion in institutional change processes through ``constitutional'' and ``collective'' layers ~\citep{frey_this_2019}. At the collective layer participants can revise the operational rules governing day-to-day activity, and at the constitutional layer they govern the structure of rule-making itself, including how decisions are made and who has a say ~\citep{frey_this_2019}.  According to our data, a key point of contention in the SE strike was exactly the lack of a formal, collaborative, and participatory process for making governance changes, i.e.  the constitutional or collective governance layers.

Research in CSCW and social computing has investigated governance crises where contributors have resorted to protest, strike, or migration. Across these studies, communities sometimes act out of grievance to protest or express frustrations to the platforms ~\cite{jemielniak_wikimedia_2016, matias_going_2016, centivany_popcorn_2016}; other times they ``migrate'' to alternative platforms ~\cite{newell_user_2016, fiesler_moving_2020}. 
As ~\citet{centivany_popcorn_2016} discusses, these alternatives reflect Albert O. Hirschman's influential ``exit, voice, and loyalty'' framework ~\citep{hirschman_exit_1970}.  According to Hirschman, people in an aggrieved relationship, such as contributors on a platform, have two possible resources: they can ``exit'' the relationship or they can express ``voice'' in an attempt to influence.  Platforms that lose a critical mass of contributors to exit become unsustainable, so they ought to listen to ``voice'' by responding and improving conditions so as to cultivate ``loyalty''  ~\citep{hirschman_exit_1970}. Notably, for ~\citet{hirschman_exit_1970} strikes are ``on the border between voice and exit'', and often undertaken when no available alternative makes exit ``credible''. A ``credible'' exit requires contributors to have somewhere alternative to go, such as a platform that is functional for them to continue their daily activities (in our case, contributing knowledge). When no such option exists, contributors' threat to leave is unlikely to carry any weight to the platforms.

~\citet{frey_effective_2023} apply Hirschman's framework to social platforms and propose ``binding mechanisms'' for ``effective voice'' in platform decision making. 
Our study contributes new knowledge of how platforms can lose loyalty of community members, and how community members facing the dilemma between exit and voice collectively organize to pursue greater influence over the platform.

\subsection{Disruptions of Generative AI to Communities and their Digital Rights}\label{background:ai}

Theoretical arguments and empirical evidence suggest that generative AI could undermine  users' motivations to contribute as well as introducing low quality content \citep{wagner_death_2025, vetter_endangered_2025}. For instance, ~\citet{burtch_consequences_2024} estimated that the deployment of generative AI technologies led to a decrease in activities on Stack Overflow (SO), particularly among new users ~\citep{burtch_consequences_2024}. 
Similarly, ~\citet{li_impacts_2024} found substantial decreases in SO answers and answer quality, particularly among heavy users who may be demotivated or lose a sense of belonging. Early evidence suggests similar effects in some types of Wikipedia articles ~\cite{lyu_wikipedia_2025, reeves_exploring_2024}.
Recent work investigating AI-generated content in Wikipedia estimated that 5\% of new English Wikipedia articles are AI-generated, and that such articles usually have lower quality such as hallucinated references ~\citep{brooks_rise_2024,mathias_undeclared_2025}. Other online communities expressed similar concerns as well. Studies of Reddit moderators show that AI-generated content decreases content quality, increases volunteer labor, and even impacts the social value of online communities~\cite{lloyd_there_2025}. Our work contributes evidence from the perspectives of moderators and users who were indeed concerned about these threats generative AI poses to SE in §\ref{finding:spam}.

In addition,  the success of generative AI has incentivized platforms such as Reddit, Twiter, and Tumblr to restrict access to content so as to monetize it. This trend has raised questions about the enclosure of the online commons ~\cite{longpre_consent_2024}. Our study contributes the perspectives of the SE community---a community built upon the free knowledge ethos---about the sharing of their data dumps.

\subsection{Empirical Setting}\label{empiricalSetting}

The Stack Exchange (SE) network includes 173 Q\&A websites (sites) such as Stack Overflow, Ask Ubuntu, and English Language \& Usage. Together, these sites attract over 100 million visitors every month. Users ask and answer questions about a wide range of topics ~\cite{se_history_24}. SE's unique community-driven model stands out among online platforms. Each SE site operates semi-autonomously with its own guidelines and rules created and enforced by moderators. For example, the \textit{Interpersonal Skills} site has strict policies on comment usage, while \textit{Science Fiction \& Fantasy} has policies about flooding the front page~\cite{mithical_answer_2020}. However, all community members must follow a universal Code of Conduct that applies across the entire network.\footnote{https://meta.stackexchange.com/conduct} Sites on
the platform use a gamified reputation system in which reputation points and achievements encourage users to ask high-quality questions, provide usable and precise answers, and participate in community governance~\citep{cavusoglu_can_2015}.

What makes SE particularly distinctive is how moderation work is distributed to enable community self-regulation and distribute moderation work~\citep{atwood_theory_2009}. As users gain reputation, they unlock escalating moderation tools—from flagging and editing posts, to closing or deleting content through consensus and monitoring otherwise private data about activity in a site.\footnote{https://meta.stackexchange.com/help/privileges} Unlike Reddit communities where existing moderators or the platform's admins grant moderator status, on SE moderation powers are comparatively open. The SE community also elects a small number of moderators to have independent authority to handle complex issues. In addition to the tiered moderation approach, SE also relies on community members to beta test the site and to operate community-developed moderation tools such as SmokeDetector, a spam detection tool. Besides the moderators and users who contributed to the network as volunteers,  official  \textit{community managers} are SE, Inc. employees who serve as mediators between the company and the user base. They listen to community concerns, support moderators, and help align company actions with community needs. Stack Exchange was thus founded under a strong ethos of participatory governance: Users are not only content contributors, but also stakeholders responsible for shaping and maintaining their space and knowledge commons.

Nevertheless, as discussed below in §\ref{finding:grievances.governance}, the platform still faces governance struggles. Following the introduction of generative AI technologies in 2022, SE moderators created new rules prohibiting AI-generated content after noticing an abrupt increase in low-quality content. Meanwhile, new contributions to Stack Overflow, the largest Stack Exchange community, sharply declined ~\citep{burtch_consequences_2024}. However, SE, Inc. subsequently released a different AI moderation policy publicly, which asked moderators to stop banning users on the ground of posting AI-generated content. At the same time, SE, Inc. also ceased publishing the ``data dump", which is the archive of SE content previously hosted by the Internet Archive, and made plans to monetize its free API service for data access. 


The strike unfolded rapidly, first led by moderators and then followed by users. Early strike calls circulated within the moderator community via channels including SE chat rooms, Discord servers, and private communication among moderators. A private Discord channel facilitated planning and communication among moderators. As moderators joined the strike and halted their moderation activities, moderation queues and spam quickly accumulated, especially on large sites such as Stack Overflow. Moderators also publicly called for users to join the strike in their strike declaration post on the Meta Stack Exchange (henceforth ``Meta SE'') site where moderators, users, and staff come together to discuss platform policies. The strike became a joint collective action among moderators, users, and former SE employees.

\section{Methods}

We conducted a qualitative study using two complementary data sources: online discussion threads including both posts and comments from Meta Stack Exchange, and semi-structured interviews with community members. We intentionally collected data from two different sources to serve distinct but connected purposes~\citep{creamer_introduction_2018}. Integrating real-time public discourse with personal retrospective interpretation in this way gives us a more comprehensive understanding of the history of Stack Exchange and the strike. Online discussions show what contributors did and said during the strike ~\citep[e.g.,][]{gupta_being_2025}, including tensions between SE, Inc. and community members' perspectives over free knowledge, and the different roles community members played in the strike. The online discussions also constitute a comprehensive body of actors' public messages and actions as they experienced the AI crisis and moderation strike in real-time. Yet online discussions only offer a partial view because users usually comment with specific goals, and platform norms in mind. As a result, discussions are a data source limited by actors' strategic purposes and situated context. To complement this, our semi-structured interviews were designed to probe deeper into these emerging themes, validate whether the public discussion aligned with participants’ accounts, and capture personal long-term reflections beyond what was visible in public discourse. Although these interviews were limited by our sampling and participants’ recollections, they provided a more complete explanations of events, messages, and what they meant. The interviews also surfaced perspectives from people holding a range of roles within the SE community, which help us understand more deeply community members’ diverse motivations and perspectives on the strike.


\subsection{Meta Stack Exchange Data}
Meta SE is a SE site specifically for network-wide policy discussions and is the primary location to observe public communications as the community mobilized to pressure SE, Inc. during the conflict. These discussions were cross-cutting and included a broad range of views about strike strategies and platform policies. 

We collected data from Meta SE including questions (posts) and comments made between June 5, 2023 and September 9, 2024. We sampled posts purposefully, by identifying strike-related posts via search for the keywords “Strike,” “Negotiation,” “AI Policy,” “Data Dump,” and “Content Moderation” ~\cite{zhangQualitativeAnalysisContent}.  We refined the dataset by applying two additional criteria: First, we included posts with at least 1 comment because we are interested in community discourse. Second, we included posts receiving at least 10 upvotes or downvotes because these garnered community attention. These steps resulted in a final dataset of 15 posts, which together contained 194 answers and 1,876 comments, yielding a total of 2,070 messages.

\subsection{Semi-structured Interviews}

We conducted interviews with 14 current and former SE users, moderators, and employees. Our interview protocol was informed by our initial analysis of Meta SE discussions. For example, SE community members expressed a range of differing opinions on Meta SE about the new AI policies and the enclosure of the data dump. We thus crafted our protocol to deepen our understanding of their attitudes about these matters. Likewise, we observed many expressions of frustration about governance and communication breakdowns in Meta SE discussions and this informed us to probe our interviewees for their perceptions of SE, Inc.’s decision-making. Our interviews with each community member opened with questions about their experience and motivations for participating in SE sites, followed by why they participated in the strike (or not) and their attitudes about the enclosure of the data dump and AI policies. We also asked participants to recall how the strike unfolded and what specific actions they took to support the strike.  

To recruit participants, we directly contacted SE moderators and users via publicly available contact information such as email addresses, Discord IDs, and LinkedIn profiles shared on their SE profile pages. Additionally, many SE moderators participated in a Discord server created during the strike. With the approval of the administrator of the SE Moderators Discord Channel, we sent a recruitment message in the server's Announcement channel. In total, we contacted 195 community members from SE and received 24 responses, with 14 semi-structured interviews conducted from 8/11/2024 to 4/03/2025. Table \ref{tab:intervieweeDemographics} provides information about interviewees' roles and years of experience on Stack Exchange. We define a user as  active if they contributed questions and answers at least once in the years they were active and as passive otherwise.
The maximum interview length is 61 minutes, while the minimum interview length is 24 minutes. The average interview length across the 14 interview participants is approximately 43 minutes. We compensated each interviewee a \$25 virtual gift card.  This study is approved by the IRB office of the first, third, and fourth author's institution. The second author only had access to anonymized data and was not involved in the recruitment and interview process.

\begin{table}[H]
    \centering
    \begin{tabular}{lll}
    \hline
         ID&  Role& Years on Stack Exchange\\
    \hline
         P1&  Active SE User& 6 years\\
         P2&  Former SE Moderator& 14 years\\
         P3&  Passive SE User& 5 years\\
         P4&  Former SE Moderator/Former Community Manager& 8 years\\
 P5& Current SE Moderator&13 years\\
 P6& Former SE Community Manager&12 years\\
         P7&  Current SE Moderator& 12 years\\
         P8&  Current SE Moderator& 9 years\\
         P9&  Former SE Staff& 13 years\\
         P10&  Former SE Moderator& 13 years\\
         P11&  Active SE User& 13 years\\
 P12& Active SE User&5 years\\
 P13& Active SE User&11 years\\
 P14& Active SE User&6 years\\
 \hline
    \end{tabular}
    \caption{Interviewee Demographics: “Years on Stack Exchange” refers to the length of time each participant had been active on their most frequently used SE site. Interviewees represented different roles on the platform, including regular users, moderators, community managers, and SE staff. Moderators are volunteers elected by the community, while community managers are employees of SE, Inc. who act as liaisons between the company and the community. The label “Former” before a title indicates that the interviewee no longer volunteers or works for SE, Inc.
}
    \label{tab:intervieweeDemographics}
\end{table}

\subsection{Data Analysis}

We analyzed our data using an inductive thematic analysis~\citep{braun_using_2006}. Open coding began with the discussion posts obtained from the API. The first author used inductive open coding such as in-vivo coding and descriptive coding of each post's answers and comments and wrote memos throughout the coding process. The first author then categorized the codes into categories and subcategories by discussing them with the second and fourth authors through an iterative process. Some emergent categories were conceptual processes rather than merely descriptive topics, such as community motivation and values~\cite{saldanaCodingManualQualitative2013}. After defining our coding schema, the second and fourth authors validated the schema by applying it to a sample of the data to ensure consistency. After several rounds of iteration, we finalized our coding schema and completed coding the entire dataset. For interview transcripts, we reused the codebook from our analysis of discussion posts but made iterative changes to capture new findings from the interviews. The first author coded all interview transcripts, with regular consultations with the second, third, and fourth authors. 

Following the coding process we identified and refined overarching themes that captured patterned meanings across interviews and discussions. We iteratively grouped related categories, compared them across participants, and connected them to our research questions. This analytic step allowed us to progress from descriptive coding to articulate higher-level themes: (1) tensions in governance, (2) contributors’ values of high-quality knowledge production, (3) tactics community members used for mobilization, and (4) alternative spaces for knowledge sharing, which we present in §\ref{sec:findings}. After arriving at these themes, we recognized that they were profoundly linked to aspects of Hirshman's exit, voice and loyalty framework~\citep{hirschman_exit_1970}, and~\citet{frey_effective_2023}'s related concept of ``effective voice'' that together serve as our analytic lens.

When reporting quotations below, we include quotations from both the online discussions and interview participants. We used P\# to refer to interview participants and Comment\# to refer to a specific comment thread of online discussion posts. For online discussions, we did not retain usernames in order to protect privacy, but all examples were drawn from publicly accessible discussions. We have lightly edited quoted comments to reduce the risks of de-anonymization. As mentioned in earlier sections, interviews provide individual perspectives, while online discussions capture real-time, community-level deliberation. Thus, we treated both interviews and online discussions as primary sources, and using both datasets allows us to triangulate the findings across complementary forms of community expression.

\section{Findings}\label{sec:findings}

We first answer RQ1 about what led to the strike with findings in terms of the long-term grievances between the community and the platform in §\ref{finding:grievances.longterm}, and in terms of emergent issues related to openness and knowledge quality that sparked the strike in §\ref{finding:grievances.strike} .  We then answer RQ2 about how the community mobilized, highlighting tiers of leadership and communication §\ref{finding:mobilize}. Finally, §\ref{alternativeKnowledgeSharing} documents the aftermath of the strike, answering RQ3. Figure \ref{fig:timeline} visualizes how the conflict unfolded, highlighting key events and concomitant exit, voice, and loyalty dynamics.


\subsection{What Led to the Strike: The Longstanding Grievances}\label{finding:grievances.longterm}

Across interviews and public discussions, participants described a long-term deterioration of their relationship with SE, Inc. that set the stage for the strike. They perceived a dramatic shift over time where the company no longer treated them as partners, as it failed to understand community needs and often acted unilaterally without consulting the community. Here, we begin to answer RQ1 by showing how years of grievances particularly over governance (§\ref{finding:grievances.governance}) and lack of transparent communication (§\ref{finding:transparency}) contributed to an acrimonious community-platform relationship and how this deteriorating relationship helped impel the community to strike. We complete our answer to RQ1 in §\ref{finding:grievances.strike} which presents our findings about the role of the AI emergency and SE, Inc.'s policies in the strike.


\subsubsection{Governance Shifted from Participatory to Unilateral}\label{finding:grievances.governance}
SE contributors valued the historically participatory nature of platform governance on SE, but noticed a shift toward unilateral decision making by SE, Inc. in recent years. SE contributors reflected on their role as stakeholders in the SE network since the beginning, as exemplified by the following comment:  

\begin{quote}
\textit{We're more than just a focus group though; we're stakeholders. While not in in terms of money, the network depends on the active support and contributions of its Community, as the sites were designed to be community-driven from the very start.
This was well understood and publicized in the early days as a feature, as I understand it.} [Comment 2]
\end{quote}

\noindent The comment notes that the company created a strong expectation in their early history that the platform would operate with the ``buy-in'' of the community in a deeper way than a mere ``focus group'' that expresses voice, but without a meaningful stake or influence. 

SE contributors thus hold a persistent expectation for participatory governance, believing their participation is not merely symbolic, but an indispensable factor in the company's success. They expected themselves to be ``\textit{actively engaging in decision-making process}'', rather than ``\textit{being defined by monetary value}'' [Comment 1]. 

However, SE, Inc.'s decision making became increasingly unilateral. Several participants and comments discussed a past watershed in SE, Inc.’s in this trend, the ``Monica incident", during which the company was seen as completely disregarding its commitments to participatory governance. In 2019, without prior warning or notification, the company revoked the moderator status of a well-known moderator known as Monica for unclear reasons. Multiple moderators resigned in protest after witnessing how Monica was treated by the company. 
This incident was widely mentioned in public discussions around the 2023 strike. In our interviews and in their public discussions, contributors pointed to this moment as a marker of the longer-term breakdown in the relationship between SE, Inc. and its contributors. For example,  P11 connected the Monica incident with ``several flare-ups since 2020'' that all led to ``a cascade of frustration and further separation between the company and community'' [P11]. For many community members, the Monica incident exemplified a broader governance problem: SE, Inc.’s governance had become increasingly unilateral, with little room for community input.

More broadly, SE contributors also found the unilateral approach to governance by SE, Inc. reflected in the platform's routine operation. Tensions around decision-making became evident when SE, Inc. attempted to implement changes against the community’s wishes. For example, interviewees thought that some UX researchers at SE, Inc. have never used the platform themselves. When SE, Inc. proposed removing the reputation requirement for voting based on the recommendation of its UX researchers—a key element of content quality control. As P4 explained, the reputation-based voting system does not only motivate users to contribute, but also facilitates finding the best answers—``instead of a forum where you read a hundred posts… it’s always the top one or two that answer your question." To community members, the reputation threshold is essential for quality control: It ensures that only experienced contributors have influence over ranking questions and removing spam and low-quality answers. SE, Inc.'s proposal to remove the voting reputation requirement was perceived by community members as SE, Inc. no longer valuing community perspectives in making decisions about the platform. 

Beyond imposing unwanted changes, contributors also described a broader pattern of neglect of community needs, as the company failed to expand infrastructure and tools to support a growing community. According to P6, a former community manager, SE, Inc. failed to invest in the community's tools as the network grew. P6’s account reflected a common frustration: while SE, Inc. continued to benefit from the labor and knowledge produced by communities, it has failed to maintain and invest in these communities, losing the community's trust and loyalty.
\begin{quote}
\textit{The community has been stuck with this system for five years. And while it may have worked 10 years ago when it was much smaller scale, now with so many more people using it, it just doesn't work and it needs to be fixed.} [P6]
\end{quote}

\subsubsection{SE, Inc. Lacks Clarity and Transparency in Communication}\label{finding:transparency}

SE contributors and moderators also attributed the strike to longstanding clarity and transparency issues in SE, Inc.'s communication. Many comments criticized SE, Inc.'s announcements as often vague about the goals and outcomes of upcoming changes.  As P1 explained, ``If the community asks something, the company will sometimes give a pretty vague answer. And when the company makes an announcement, they are often very vague about what they want, why they are making this change, and so on." Some also found SE, Inc. unresponsive to community questions and commented that ``communication was always one-sided, with SE, Inc. resistant to receiving feedback from the community'' [Comment 4]. The lack of clarity and transparency in SE, Inc.'s communication is especially a concern for community managers, who are hired to mediate between SE, Inc. and contributors. The two participants we interviewed who were community managers recounted how they were ill-equipped to fulfill their responsibilities due to the lack of clarity and transparency in SE, Inc.'s public communication. According to P6, a community manager and long-time contributor, the company that had once been transparent became increasingly opaque. Although P6 valued their role as a mediator, they recalled becoming disoriented about what information could be shared, which ultimately led them to ``stop talking'' and left their role as mediator meaningless.
\begin{quote}
\textit{
[SE, Inc.] shut off all public communication about business plans they wanted to make privately \dots the only people they talked to were under NDA \dots
I got to the point where I didn't actually know what I was allowed to talk about and what I wasn't allowed to talk about. And so I just stopped talking. 
} [P6]
\end{quote}

Many contributors echoed P6’s frustration, framing lack of transparency as a recurring grievance. Without clear communication, they felt excluded from governance and anxious about the company’s direction.

\subsubsection{A History of Grievances Set the Stage for Conflict}
Participants attributed the strike to the accumulation of tension between the company and the community over the years, and when disruptions from AI technologies affected the community, contributors were distrustful of SE, Inc. P11 emphasized that the strike was the community's last resort following repeated conflict episodes—most notably the Monica incident—which had already pushed many contributors to the fringe of disengagement. According to P11, each breach of the company's past commitments to openness, transparency, and community-centered governance, only “served as fuel” for growing frustration. By the time SE, Inc. announced its AI policies, contributors felt that the company had ``painted itself into a corner", leaving moderators and users exhausted and disappointed. As P11 explained, by the time the strike began, ``we just had enough—we were just fed up with it."



Several others similarly pointed out that while SE, Inc.’s new policies on moderating AI-generated content and restricting data dump access were the immediate catalysts, the deeper issue was the community has grown skeptical of the company's receptiveness of community voices. This framing echoed public statements made by moderators during the strike, including Meta SE posts clarifying that the protest was not motivated by ``a general hatred towards AI,'' but by the company enforcing policy changes without meaningful community input.
\begin{quote}
\textit{So if the AI problem had happened in 2015, I don't think a strike would have occurred. It might have caused some issues, but things would have moved on. But when this happened, we were already coming from Monica's case and other incidents where people had started losing trust in the company. } [P10]
\end{quote}
As P10 explained, in earlier times, conflicts might have resolved easily, but  due to repeated neglect of their voices, many participants felt left little hope that SE, Inc. would listen during the AI emergency. In their accounts, the company's past violations of the community's expectations damaged the community's loyalty to the company, increasing the chance that the community would turn to exit as a recourse. The AI policies were the tipping point. 

In sum, participants framed the AI emergency as the spark that ignited community's frustrations, not the sole cause of the strike. While bystanders might interpret the community strike as simple opposition to AI, contributors consistently described it as rooted in years of a deteriorating relationship with the company.

\subsubsection{The SE Community Demands Participatory Governance}\label{finding:grievances.values.beg}
During the strike, contributors and moderators made explicit demands for stronger mechanisms for authority and accountability in platform governance, based on the longstanding grievances they have had with the platform. Specifically, they urged SE, Inc. to reestablish trust by gathering feedback before implementing platform-wide changes, seeking to ensure their voice would be effective. As one contributor stated, ``before making major policy or software changes, Stack Exchange, Inc. must engage with the community, gather feedback, and take that input into account,'' because the company had ``kept implementing changes that harmed the community and conflicted with the network’s goal of sharing knowledge'' [Comment 6].

Community members also proposed governance mechanisms to institutionalize their demands in concrete and enforceable ways.  For example, one moderator recommended that 
\begin{quote}
\textit{Any proposed mandatory policy affecting the entire network must go through a period of feedback and review by moderators for at least seven business days, so moderators can provide feedback and have the policy revised before it takes effect. }[Comment 7]
\end{quote}
What mattered most to the community was ensuring that their voice in decision-making was not merely symbolic but effective for shaping outcomes. Proposals such as mandatory feedback periods reflected a desire to institutionalize accountability for the platform so that contributor voices would translate into meaningful change. 

\subsection{What Led to the Strike: The Emergent Grievances}\label{finding:grievances.strike}

We found in §\ref{finding:grievances.longterm} that long-running grievances about unilateral governance and a lack of  transparency made the community distrustful, frustrated, and anxious about SE, Inc. and its response to emergent threats from AI.  While this dynamic was surely a factor leading to the strike, the community also recognized that AI technologies prompted an emergency that proximally caused the strike. Below, we describe how SE, Inc. failed to support the community's core values of quality, free knowledge, and attribution in the wake of AI-led disruptions.  


\subsubsection{Low-Quality AI-Generated Content Floods the Site}\label{finding:spam}

One emergent reason for the strike is the community's concerns about the influx of low-quality AI-generated content. Creating and maintaining high-quality content is one of the most important values for Stack Exchange contributors who are motivated to create a useful, accurate, and well-organized public knowledge base. Many interviewees said that this focus on quality is what makes contributing to the community engaging and meaningful.

As several interviewees explained, the high visibility of Stack Exchange content through search engines motivated them to contribute high-quality knowledge. P1 described ``just typing a question in Google" and finding that ``Stack Overflow... has a lot of useful resources" in an accessible Q\&A format. P3 similarly noted that ``Stack Exchange is one of the top results" when searching for information. For contributors, this visibility strengthens the community's purpose of maintaining quality, and motivated them to take part in improving the content. P12, a long-term user with some moderating powers, emphasized SE as a place for ``enthusiasm, quality, and subject matter expertise," where the goal was to help future visitors access high quality answers rather than just to resolve a question in general.

From the contributors’ perspective, AI-generated content threatened the community's value of contributing high quality knowledge. Following the release of large language models such as ChatGPT, many newcomers began submitting AI-generated answers to quickly gain reputation points. These answers were often posted with minimal verification and described by long-term contributors as ``spam" or ``copy-and-paste abuse." Although LLMs could sometimes generate correct answers, contributors stressed that LLM outputs are unreliable. 

Moreover, contributors also worried that AI-generated spam would pollute not only SE’s content but the wider information ecosystem. P2 noted that generative AI “hallucinating too much makes it harder for actual, accurate information to be seen,” while P9 warned of a vicious cycle in which “clueless consumers begin to regurgitate remixed prompt outputs back into the space that AI trains from.” For contributors, maintaining SE’s quality was thus both a community value and a public service, protecting information seekers from the noise of inaccurate AI content.

This tension was intensified by SE, Inc.'s sudden announcement that asked moderators to stop suspending users for posting what they suspect to be generative AI content. Moderators saw the announcement as SE, Inc. ``blaming moderators for relying on AI detectors to moderate AI content'' and for driving away newcomers who used AI tools to write answers. They shared in discussion posts that their moderation decisions were primarily based on heuristics to ensure the quality of knowledge in their communities. SE, Inc.'s sudden demand startled moderators and undermined their goal for maintaining knowledge quality. 


Additionally, moderators felt that the labor they put into ensuring knowledge quality went underappreciated and unrecognized by the platform. One moderator described their work as ``sorting the junk [low quality, AI generated content] out'' in comment. P10, a former moderator who volunteered for 13 years, explained that ``high-quality knowledge production depends on contributors’ unpaid labor".  While the rise of AI spam created a surging increase in their workload, SE, Inc.'s announcement about moderation policies for AI-generated content was perceived as devaluing the labor of moderators upholding their community's quality standards.




\subsubsection{Shutting Down the Data Dump Violated the Free Knowledge Ethos}\label{finding:accessibility}

Many contributors view SE as a collaborative, long-term project to produce, curate, and share free knowledge globally. Our interviewees valued \emph{accessible distribution} of knowledge to everyone who might need or reuse it.  Because the data dump was instrumental to the SE community’s shared free-knowledge ethos, the company’s surprising decision in March 2023 to pause its release was met with community outcries. Even though similar delays or issues have occurred in the past, this incident, coupled with the contemporaneous announcement of SE, Inc.’s  plan to monetize its API service provoked public outrage. The community grew concerned that the nominally temporary pause in the publishing of data dumps would escalate to permanently restricted data access.  For example, our interviewees brought up examples of projects that were possible thanks to the data dumps, such as Kiwix and Brent Ozar’s Unlimited SQL which provide offline access to educational content. Similarly, the ``Overflow Offline Project,” an offline version of SE knowledge for people in geographically isolated areas, such as Antarctica, or without Internet access, such as prisoners ~\cite{so_offline_22}, was also mentioned in public discussion.  SE contributors expressed their concern that these collaborative projects would also be impacted by data dump restrictions. 

Although SE data could be accessed through the Data Explorer (SEDE) and API, these downstream applications mentioned depended on bulk data access afforded by the data dumps. Moreover, the change violated contributors' expectations that the knowledge they contributed to SE would remain open and accessible for the general public, as an interviewee explained:
\begin{quote}
\textit{A lot of people were understanding of the idea of monetizing the commercial aspect. That wasn't a problem, but violating the principle of sharing this data to researchers and people who might use it in the future, that was a big violation.} [P4]
\end{quote}
\noindent P4 draws a distinction between monetization and the open knowledge ethos. While many contributors understood that SE, Inc. might seek commercial use of the data, they did not object to monetization \textit{per se}. The concern, instead, was that stopping the data dump violated the principles of open knowledge sharing and reuse.


\subsubsection{AI Poses Attribution Issues}\label{finding:attribution}

Contributors repeatedly raised concerns that generative AI’s use of their contributions violated not only the CC-BY-SA license’s attribution requirement~\cite{license_ccbysa_24}, but also the recognition and credit that motivated them to contribute in the first place. They expected AI companies and other external users to respect the CC-BY-SA license applied to the user generated content of SE network:
\begin{quote}
\textit{I assume some of us maybe understand that our work benefits others, even corporations...but that was always supposed to come with attribution. Has Stack Exchange, the company, considered that there may be class action exposure for violating millions of users' content licensing? }[Comment 10]
\end{quote}
\noindent Comment 10, like many other contributors, expressed concern that if their knowledge were used for AI training, there would be no proper attribution to the contributors, thereby breaching the CC-BY-SA license. 

P10 also emphasized that receiving appreciation messages would provide a sense of fulfillment: ``We know we were creating a better place but also getting some acknowledgment '' [P10]. However, the reuse of SE content for AI training fails to acknowledge the community's contributions:  
\begin{quote}
  \textit{We have been providing a lot of data for 15 years now, and now we are not even mentioned as in credit about having being the creators of that information. } [P10]
\end{quote}
Taken together, contributors saw critical attribution issues in how generative AI presented their knowledge and challenged SE, Inc.'s decision to support such content reuse.  

\subsection{How Community Members Mobilize: Building Collective Power and Withholding Volunteer Labor}\label{finding:mobilize}

We now turn to answer our RQ2 about how community members mobilized collective power.  
We found that the SE community used a tiered structure of organizational communication, beginning with a small, enclosed group of moderators and then spreading across the users on the network. We unpack this communication process in §\ref{finding:tiered.communication}. Members of SE participated in the strike by withholding contributions and moderation, as discussed in §\ref{finding:withholding.contribution}. 

\subsubsection{The SE Community Organized Tiered Collective Action} \label{finding:tiered.communication}

The strike did not emerge spontaneously, but rather required active coordination and communication.  
From our interview and online discussion data, we found that Stack Exchange moderators initiated the strike as early whistle-blowers and information disseminators. According to P8, a moderator who extensively facilitated the strike, moderators initially convened to coordinate and communicate on a private Discord channel which was conceived as a small and private command center for focused and efficient communication. Within this channel, moderators discussed strategies for resisting SE, Inc. confidentially among themselves. They later elected representatives to engage in formal negotiations with SE, Inc.'s leadership.

Moderators came to understand that support from the broader SE user population would increase their leverage. They used communication channels off of SE to reach and communicate with these users, such as a public-facing Discord space and a website hosting the open letter as well as frequent updates about strike and negotiation progress on a post on Meta Stack Exchange. The information they published to these channels cascaded to other online communities and social networks, creating widespread awareness of the strike. One of our interviewees who is primarily a user, not a moderator, learned about the strike through such publicly available discussions:
\begin{quote}
\textit{I did see when the posts were on Meta and I followed the links of related posts to see like, oh, this is what happened, this is what happened. Going back further, there's a bunch of stuff about Monica that I've read about and just generally learning about this is what has happened in the past, and seeing the letter, I support them.} [P3]
\end{quote}
Although a SE contributor, P3 only learned about SE's troubled history after becoming aware of the strike from posts on the Meta Stack Exchange site. Learning of the Monica incident influenced them to support the strike action. Thus, calling for broader participation also spread information about the company's past missteps. 


\subsubsection{The SE Community Withheld Contributions and Moderation} \label{finding:withholding.contribution}

The SE community selected a strike strategy aimed at halting both moderation work and knowledge contributions in order to compel the company to recognize the crisis and negotiate. P5 notes that withholding moderation labor from a large, high-traffic site like Stack Overflow is dramatically disruptive.  


\begin{quote}
\textit{If all the moderators stopped work for five days, the work wouldn't get out of hand as much as it would on the biggest site, which is Stack Overflow. So if the Stack Overflow moderators decide to stop working, all the queues would fill up immediately. So they have a little bit more leverage.} [P5]
\end{quote}
\noindent A temporary halt in moderation might be manageable on smaller sites, but on Stack Overflow, even a short break rapidly overflows moderation queues and the consequences of inaction are more immediately felt. This illustrates how a strike's potential to exert power depends not only on the quantity of participants, but also on their position within the community.


The strike ultimately involved cross-role collaboration among moderators, users, and former SE employees. many users joined the strike by suspending their activities on the SE network, including asking questions, providing answers, and voting. As they gathered in Discord as well as other spaces on SE sites and across the web, they organized their activities from bottom-up, with loose and decentralized coordination. These efforts resulted in a sustained strike that ultimately pressured the company to negotiate with community representatives.

Participants acknowledged that the strike created an opportunity for the community and the company to engage in open discussions about the platform's future. In this sense, the strike was generative, creating a space for all parties to identify opportunities for participatory governance through dialogue. As one participant explained:
\begin{quote}
\textit{I think that one of the main things that the strike achieved was getting an avenue to open discussions so that more investigation could be done.} [P6]
\end{quote}
As P6 suggests, strikes catalyzed discussions and created spaces where the community can continue to organize and plan longer-term reforms. Rather than treating the protest as a one-off conflict, SE, Inc. acknowledged the need for ongoing consultation, thereby granting the community a more binding role in governance. This development reflects~\citet{frey_effective_2023}'s concept of effective voice: The main purpose of the strike was to help the community hold administrators accountable and substantively influence them.


\subsection{What Happened in the Aftermath: Migration and Preservation}\label{alternativeKnowledgeSharing}

We turn now to our findings to answer our RQ3 on the aftermath of the strike. We found SE community members explored alternative ways to contribute free knowledge independently of SE, Inc. They migrated to alternative Q\&A platforms (§\ref{migration}) and advocated for decentralized knowledge sharing and archival activism to preserve the knowledge they had created (§\ref{archival.activism}).

\subsubsection{The SE Community Seeks Alternatives through Migration} \label{migration}

Although the strike resolved with an agreement giving the community a larger role in governance as discussed in §\ref{finding:withholding.contribution}, several contributors indicated in their interviews and public comments that they still distrusted SE, Inc.  Given the history of conflict with SE, Inc., these contributors sought alternative platforms that can support participatory governance. Some of our interviewees described how during and after the strike they contributed less to SE and more to their personal websites, blogs, and other peer production projects. The most prominent alternative attracting a significant number of SE contributors, however, may be Codidact, a Q\&A platform similar in design to SE, but structured as a Community Interest Company (CIC), where a non-profit organization and operates in the community's interest.

The migration to Codidact began after the Monica incident in 2019 and remained an ongoing effort. Some contributors who migrated in 2019 became dual users of both sites for a time, and it was not until the 2023 strike that they fully exited SE. Such partial migrants appended ``moved to Codidact'' to their usernames, both publicly signifying their departure and advertising Codidact as an alternative. During the strike, these users also frequently replied to posts on Meta SE related to grievances about SE, Inc. For example, one commenter noted that ``the way the Codidact Foundation is set up satisfies most, if not all, of what you ask for here.'' [Comment 13] Some also migrated to TopAnswers, a similar but smaller platform.

According to our interviewees and the public discussion, Codidact offers contributors full rights over their content and a participatory model for governance. Community feedback is actively gathered through meetings and discord channels to guide platform improvements, and technical measures such as robots.txt are used to inform bots about what is considered unauthorized scraping. These features demonstrate Codidact’s alignment with SE contributors' goals and wishes, especially around participatry governance. 

However, Codidact faces its own challenges. Participants acknowledged that its dependence on volunteers in various timezones leads to difficulties in coordinating work on technology development, workflow, and effective communication. 
\begin{quote}
\textit{Mainly because of my own work commitments and some communication issues (e.g., time difference, non-overlapping volunteer shifts) with the Codidact team. I wasn't really able to complete [a development task]}.~[P12]
\end{quote}
\noindent P12’s comment reveals the practical challenges in mobilizing around exit strategies.  Despite the motivation to exit, work to build an independent platform such as Codidact is constrained by work obligations, time zone differences, and coordination issues. Exit thus requires more than ideological alignment; it also demands time, infrastructure, and collective coordination.

Even with coordination among volunteers, Codidact still faces additional challenges, such as technical and design limitations. For example, some interviewees thought Codidact's UI and interface ``\textit{are quite outdated at the moment}" [P12]. 
Additionally, P2 raised the limitation of robots.txt file. Namely, the settings are able to “block the ethical crawlers, block the crawlers from the ethical companies,” but “there’s nothing you can do about the unethical companies.” Even though robots.txt can block companies that voluntary comply with the standard, increasingly many crawlers ignore it.

Finally, although many moderators migrated based on strong social ties, other long-term SE users were hesitant to leave as the following comment exemplifies:
\begin{quote}
\textit{Currently, Stack Exchange is useful because it provides value I can't get somewhere else. While Codidact is great as well, it is quite small, and there is really nowhere else where there are humans answering questions in a formal, curated, high-quality manner. That's what makes this place unique, that's the reason I'm here. }[Comment 11]
\end{quote}
\noindent This contributor is aware of the alternative platform but does not find it an adequate substitute for SE.  They describe SE's advantages as having a large base of contributors and knowledge built over the past 17 years. This resonates with ~\citet{frey_effective_2023}'s argument that even when exit seems easy, network effects may help entrench the previous platform's  dominance.

Those who migrated to Codidact also recognized that migration or exit, while possible in theory, is challenging. Coordinating a migration requires more than simply creating a technical alternative; it also depends on sustained volunteer labor, shared norms, and mechanisms to bring an entire community along. Without these supports, migration efforts risk fragmenting one community into smaller groups. As P2 put, ``We set up a site, but as it turns out, communities don't migrate. Communities fragment.'' The Codidact case illustrates that exit requires not only volunteer work,  but also large-scale coordination and network effects that allow a community to reproduce itself elsewhere. 


\subsubsection{The SE Community Shares Decentralized Knowledge Through Archival Activism}\label{archival.activism}

As mentioned above in §\ref{finding:accessibility}, the free knowledge ethos is widely shared among SE contributors and moderators and is reflected in their efforts to share their knowledge through archival activism after the strike. 
Fearful that SE, Inc. could limit access to data dumps in the future, contributors explored different ways to protect the community’s knowledge base, a form of archival activism. Some proactively seeded torrent files of the dumps, to create a distributed network of backups. This peer-to-peer approach improves the availability of historical data, even if official channels have become restricted. As told by one comment:
\begin{quote}
\textit{The data dump isn’t gone yet, but I don’t believe this will be the last time SE, Inc. tries to undermine it. If you want to support the community in pushing back, there are a few ways to start: contribute code, speak out against SE, Inc.’s actions that hurt the community, and help seed the data dump torrents. By seeding, we make it much harder for the dumps to disappear quietly, even if they’re removed from archive.org. If you can, consider setting up a seedbox—it’s one of the most reliable long-term ways to keep this data from being lost forever. }[Comment 12]
\end{quote}
\noindent Comment 12 frames the company’s restriction of data dump as an ``attack'' that contributors ought to resist not only by expressing voice, but also by seeding torrents which might support future exit. 
Together, these decentralized and archival practices demonstrate the SE community's commitment to knowledge as a public good. By ceasing dependency on a single platform, taking active roles in preserving community-created content for future learners, researchers, and developers,  these actions again showcase the SE community's devotion to the free knowledge ethos.  


\section{Discussion}

We studied the 2023 Stack Exchange (SE) strike with data from public discussions and private interviews to investigate why SE moderators and contributors went on strike, how they organized, and what they did in the aftermath. Our thematic analysis revealed that although AI technologies caused emergent problems for the SE community as well as broader anxieties about AI, neither these problems nor distaste for AI brought the community to strike. Rather, the strike happened because these problems were inextricable from SE, Inc.'s long-term neglect of the community's voice, despite SE's affordances for community-driven moderation and participatory governance. In the terms of \citet{hirschman_exit_1970}, by not giving meaning to the community's voice, the company gradually lost the community's loyalty, increasing their pressures to exit. In §\ref{discussion:effective.voice} we draw on these findings to discuss the prospect for platforms to use effective voice to promote community loyalty. 

Emergent problems related to AI of quality, pollution, and attribution arose in 2023, as we document in §\ref{finding:grievances.strike}. In response, the SE community used its levers of participatory governance mechanisms to prohibit generative AI and expressed voice to SE, Inc. Again, the company failed to listen, even aggravating the community's crisis with new policies. Without either meaningful voice or credible exit, SE contributors chose to strike. The SE strike was organized initially by core leaders who mobilized support from among other SE contributors and across the Internet. Based on these findings, we recommend in §\ref{discussion:online.communities} that communities can better position themselves for such crises and conflicts with platforms by cultivating alternative platforms and communication channels to prepare for collective action and exit.

Following the strike, SE community members continued to work to create and preserve free knowledge by doing so on other sites and working to preserve SE's knowledge base. Based on these findings and on the particular threats that AI presents to free knowledge communities raised in §\ref{finding:attribution} and elsewhere, we suggest future work to design attribution for generative AI in §\ref{discussion:designers} and propose a model of community-driven data governance to give online communities effective voice in matters of data reuse in §\ref{discussion:data.governance}. Finally in §\ref{discussion:tension}, we reflect on the ubiquity of community-platform conflict in light of our findings. 


\subsection{Recommendations for Online Communities}\label{discussion:online.communities}

We found that the SE community struggled to influence SE, Inc. despite the platform's relatively participatory model for governance. Some community members began building and migrating to Codidact in 2019. However, as ~\citet{frey_effective_2023} suggests, ``exit'' is rarely easy. Consistent with prior work on fan fiction communities ~\cite{fiesler_moving_2020}, full migration is often difficult: Resources and past contributions usually remain in the original platform, moving an entire community is difficult to coordinate, and as suggested by our participants, could lead to fragmented communities. This suggests that even in times of harmonious relationship with their platforms, communities ought to prepare for potential migration. A time may come when unilateral changes to platform policies contradict community values, and the community will wish to swiftly exit to an alternative.  Moreover, ~\citet{hirschman_exit_1970} theorized that ``credible exit'' would compel organizations to listen to voice. This suggests that steps to ease migration may encourage platform administrators to take community voices into account. Below, we raise examples of specific challenges that make exit difficult in practice and discuss how online communities can prepare for these challenges.

Our study raises challenges in making exit credible that communities should prepare themselves to overcome in advance of crises. Although SE supported transporting data from one platform to another, Codidact chose not to import data from SE at a large scale due to resource limitations, such as limited volunteer capacity. Moreover, the SE community struggled to coordinate migration and avoid fragmentation as contributors explored different alternative sites. In other words, the resource and volunteer labor demand of preparing for a potential conflicts with the platform are not evenly shared, with only a small group of moderators and highly active users undertaking most of the coordination, communication, and strategic planning work. 

We suggest that communities can prepare to surmount such challenges by building and inhabiting off-platform spaces for community connection, communication, and mobilization.  We observed that during the strike and the negotiation, moderators and contributors relied on Discord as an important alternative space for mobilization and deliberation, beyond the official communication channels that are only shared by moderators. Going further, holding regular community events in such alternative spaces may facilitate community-building in preparation to coordinate during potential conflicts or community migrations and by developing exit options. Social movement research shows that organizational structures created during protest often continue to sustain collective action during period of lower mobilization or reduced political opportunity ~\cite{tarrow_power_1998}. Having alternative spaces and platforms not only available, but already active in advance of crises will ease migration and make exit credible. Notably, given the widespread pattern of conflict between communities and platforms on which §\ref{discussion:tension} elaborates, communities should consider the trade-offs between using a popular and familiar commercial platform such as Discord and using one over which they have greater control, such as a self-hosted discourse forum. While commercial platforms offer convenience, they give communities little control over data or governance; self-hosting, on the other hand, provides that control but requires technical and organizational resources to maintain.


Also crucial in preparation is the institution of channels and processes for coordinating migration.~\citet{matias_civic_2019} finds that such civic labor—the routine work that keeps a community connected and organized—can help communities respond more effectively during crises. In our case, this civic labor was organized in a tiered structure during the strike. A small group of moderators initiated the strike mobilization and devoted themselves to coordinating collective action. They called for participation first from moderators and later broadcast their message to users from across the Web.  This layered, onion-like structure with core members such as moderators in the inner layers with newcomers in the outer layers resembles the pattern of legitimate peripheral participation characteristic of online communities and open source contributors ~\cite{lave_situated_1991, bryant_becoming_2005, jergensen_onion_2011}. Thus we expect that organized migrations may follow a similar structure. This tiered process presented a trade-off: While it afforded rapid, productive conversations and coordination during times of crisis, important contextual knowledge, such as about the Monica incident, tended to circulate in the inner circle before reaching the outer layers. To mitigate this, we suggest that core community members should use alternative spaces to establish communication channels for long-term community engagement, such as newsletters, websites, or blogs with which to disseminate their messages toward the periphery. By doing so, online communities can gather more diverse representation of the volunteer group, instead of relying on core members alone to organize collective action. 

Preparing for exit requires communities to routinely create data back ups. They might designate a data steward responsible for collecting and archiving community data through APIs or scraping, and uploading it to public repositories like the Internet Archive or distributing it via torrents. Not all platforms provide access to data dumps; in fact, most online communities are on platforms such as Reddit and Facebook that do not. For such communities, aggregating their shared knowledge is more challenging. One potential workaround may be data donation tools that could allow members of such communities to voluntarily share knowledge with their communities and support their goals for preservation and free knowledge, without depending on data infrastructures owned by platform operators \citep{zannettou_analyzing_2024}.

\subsection{Recommendations for Platform Operators}\label{discussion:effective.voice}

As Odysseus bound himself to the mast so as to safely hear the Sirens' song, operators of online community platforms such as SE, Inc. ought to bind themselves to listen to their communities; to give them effective voice ~\cite[][pg.127]{gowder_networked_2023,frey_effective_2023, elster_ulysses_2000}. Effective voice goes beyond feedback-gathering exercises, such as surveys or structured feedback sessions, which can gather community members' views and insights but are not necessarily binding. By contrast, \citet{frey_effective_2023} discuss a wide range of effective voice mechanisms all of which obligate platform administrators to community accountability, authority, or influence in various ways. 
Here we highlight the need for mechanisms that formalize collaborative decision making and are sufficiently binding to be enforced during a crisis. In order to better support volunteer labor and preserve the community’s values, we call for stronger community voice in shaping platform-wide policies. For example, a community representative on a platform's board having veto power over platform-wide content moderation policy changes could have forced SE, Inc. to take the community's needs into account when crafting its AI policy. Such structures seem critical for platforms to sustain a harmonious relationship with their communities in turbulent times. 

Doing so is not just for maintaining a harmonious relationship between platform operators and the communities, but also for recognizing volunteer labor. Platforms such as SE, Inc. depend on high-quality knowledge contributions and ongoing volunteer moderation work that sustain their core product and revenue model, and all these comes from the community. This creates a relationship of mutual dependence: platforms rely on communities for content and curation, while communities rely on platforms for infrastructure and visibility. Yet platform operators often encounter short-term incentives that push them to make decisions that are different from community values \citep{gowder_networked_2023}. These shortsighted decisions may help the company survive in the immediate term, but they neglect the contributors whose voluntary labor makes the platform operational in the first place. Sustaining this mutual dependence therefore requires governance structures that protect the community's long-term interests~\citep{gowder_networked_2023}.

While our suggestions for instituting effective voice discussed above promise to support a harmonious relationship, they will be strengthened by possibility of credible exit, when users have feasible alternative space to which they can migrate ~\cite{frey_effective_2021}.  Credible exit not only empowers the communities, but also reduces long-term governance risks for platform operators. Conflict is expensive for companies, and maintaining loyalty requires long-term investment in moderation, communication, and policy enforcement. Conversely, when users have alternatives, the exit process can function as a self-binding mechanism that mediates platform behavior and mitigates costly disputes.

Emerging decentralized platforms illustrate why implementing credible exit into platforms' business strategy is promising. BlueSky, for example, has explicitly marked credible exit as a design goal supported not only by making data transportable, but also by publishing design specifications and releasing open source licensed code~\citep{kleppmann_bluesky_2024}.  Design mechanisms that ensure interoperability such as portable data and accessible archives can support communities' work to build functional alternatives~\citep{doctorow_how_2022}. Protocol-based designs also promise to externalize the political and operational work of governance previously concentrated within companies so they can focus on providing services such as hosting or premium features rather than carrying the full cost of managing community conflict. In this way, interoperability and portable data do more than enhance users’ autonomy; they also create incentives for platforms to remain aligned with community values rather than short-term shareholder pressures.


In using online contributors' knowledge to train AI models, platforms ought to consider how to use the data fairly and transparently.
Across our interviews, many participants described themselves as stakeholders, not in the definition of financial ownership, but in-terms of participatory governance of the content. Thus, platforms should foster a more reciprocal and ethical ecosystem between AI developers and knowledge contributors. 
We elaborate on recommendations that can give communities effective voice in data governance matters in §\ref{discussion:data.governance}. Here we note that platforms can preserve the openness of data dumps for public use and also sell a value-added commercial version. For example, platforms may consider only charging a fee for API data access to enterprise buyers who seek a substantially higher access rate than researchers and independent developers.  For example, the Wikimedia Enterprise API provides Wikipedia's CC-BY-SA licensed content in a structured format optimized for AI workflows at a lower rate-limit than the free API.  Such a model could have afforded SE, Inc. a revenue stream  without retrenching their original commitment to knowledge sharing. We recommend that such approaches happen with community input. For instance, community input may be important for decision making about usage conditions such as mandatory attributions, financial transparency, and licensing terms. Moreover, scholars of cooperative online communities note that online platforms frequently depend on intensive volunteer labor, yet lack formal mechanisms for shared ownership or governance ~\cite{mannan_cooperative_2024}. Notably, communities also have an interest in such revenues to the extent they will be invested to benefit them such as through better moderation tools, research and development, or compensation. By sustaining a cooperative and reciprocal relationship, platforms can continue to rely on the community knowledge, feedback, and labor that are indispensable for innovation in the age of AI.


\subsection{Recommendations for Designers and Developers}\label{discussion:designers}
Our findings on community members' concerns around generative AI models and attribution issues highlight the importance of taking communities' perspectives into account when developing novel technologies. Of particular concern to SE contributors was how AI-generated information often did not properly attribute the community or contributors. We point to the need to design and build transparency mechanisms that make visible how community-produced data are used in innovative information technologies. For example, visualizations such as the Web Organizer highlight the importance and impact of user-generated content and promote understanding of how communities contribute to broader information and knowledge ecosystems~\citep{Wettig_2025_Organize}. Such honor and recognition not only highlights a community's otherwise obscure achievements and contributions, but also helps incentivize contributions~\cite{Vines_Configuration_2013}.  It is particularly crucial that developers of downstream information technologies be mindful of the ethical implications of using communities' data dumps. For example, they ought to comply with attribution norms when community members have such expectations, as on SE. 

Moreover, designers should support features that facilitate each community’s autonomy in setting AI-generated content rules, recognizing the context-sensitive nature of moderation. Since different communities hold different norms around acceptable uses of generative AI such as translation, while some prohibit it entirely. To respect this diversity, designers should enable moderators to define local AI generated content policies, including customization forms of AI involvement and disclosure requirements that aligned with community norms. Making these rules visible at the point of posting increases compliance, helping contributors understand and respect local expectations \cite{matias_preventing_2019}. Additionally, AI developers ought to carefully attend to how their products impact downstream communities and invest in mitigating potential negative impacts, such as spamming of low quality content as on SE. Failure to address these issues may lead to broader backlash from communities and potentially threaten the sustainability of the information ecosystem where AI technologies depend on the knowledge online communities produce. 



\subsection{Mechanisms for Community-Driven Data Governance}\label{discussion:data.governance}

This work surfaces problems that suggest a need for community-driven data governance models that prioritize input from data contributors. Currently, hosting platforms such as Huggingface and dataset curators predominately govern datasets, often without any feedback from the people who contributed the originally. Scholars have identified several ethical and legal risks in these AI data governance approaches, such as loss of agency for contributors ~\cite{ajmani2024data}, lack of clear licensing enforcement ~\cite{longpre2023data}, and insufficient consent from creators ~\cite{longpre2024data,ajmaniSystematicReviewEthics2023}. Our study contributes a deep analysis of a case that renders these concerns concrete and urgent. If platforms and AI developers fail to mitigate data contributors' concerns about the governance of their collective datasets, more community strikes and unrests may occur, leading to the deterioration of the online commons for all. 

Based on our findings, we propose potential mechanisms for community-driven data governance.
First and foremost, data governance must be transparent and include a community role in decision making as evidenced by the way SE, Inc.'s process to restrict data access was criticized by the community members for lacking transparency and community input. Future work may explore different transparency mechanisms (e.g. who is using the community's data, how, and for what purposes) and consensus building activities such as those applied in citizen science ~\cite{sharma_consensus_2022} to institute effective voice in community-driven data governance. In the case of SE, communities could establish a community review board, composed of users, moderators, and platform staff (such as liaisons or community managers), to actively review the data shared with AI companies. This board would oversee data reuse, provide permissions where necessary, and operate similarly to an institutional review board (IRB), following transparency and shared decision making practices. 


Given that attribution is a priority for many SE community members, it will likely be a goal for community-driven governance in cases where data contributors care about credit and recognition. Moreover, supporting attribution has the potential to improve data provenance and documentation for LLM training. Gebru et al. suggest that tools like datasheets can help track data provenance and reduce misuse of datasets ~\cite{gebru2021datasheets}. Longpre et al. has highlighted the lack of provenance information in prominent datasets makes it difficult for AI developers to verify the origins and contexts of a dataset ~\cite{longpre2024data,longpre2023data}. Given attribution's importance to data contributors and downstream technologies, platforms and communities may create terms and clauses that mandate AI developers to give proper attribution to original data contributors. 


\subsection{Broadening the Scope of Community Governance}\label{discussion:tension}

Conflicts similar to the strike we studied recur across platforms and time, and our study contributes a new understanding of why this pattern keeps happening---the lack of effective voice that empowers communities to play a participatory role in platform governance. Put another way, there exists tension between the top-down platform governance, which is often taken as the default due to company's ownership and control of platform infrastructure, and the bottom-up participatory governance that communities seek. Prior research on community governance often focused on what \citet{frey_this_2019}  called ``operational'' and ``collective'' rules---rules that define how a system operates with what roles, such as moderation roles and what community norms. Our work complicates the picture by revealing a deeper struggle experienced by community members that motivated the strike, the lack of abilities to gain decision making power to shape the platform. 

Moreover, our findings about community members' desire to be stakeholders in the decision making process reflects a fundamental power imbalance between platforms and communities that warrants further investigation. Future work should view such conflicts with a broader lens of community governance, paying attention to empowering community members in the decision making process concerning the platform. Our work suggests that pursuing effective voice---binding platforms into an agreement---may be a particularly promising approach. As ~\citet{frey_effective_2023} suggests, ``mak[ing] a community’s rules binding on administrators'' creates accountability mechanisms for administrators. The strike participants' negotiation achieved just that, formalizing agreements that SE, Inc. must take community feedback into account for future platform changes.

Finally, our findings suggest that an important question for future research is: How might the growing use of community-produced data to train AI systems---especially when these systems increase moderation workload, and decrease the long-term motivation of volunteer moderators? While moderators in our study expressed clear concerns about attribution, data use, and quality of AI-generated content, the long-term implications for sustaining volunteer labor remain uncertain in the age of AI. We view this as an important question for future research, particularly as online communities continue to depend on civic labor that may be constrained by AI-related governance burdens. Future studies could explore how generative AI reshapes practices of content moderation, peer production, and community governance in general. For example, future work could quantify how much volunteer labor is shifted to identifying and examining AI-generated content---a task that moderators in our study described as already becoming burdensome.

\section{Limitations}
We sought to produce a comprehensive understanding of the strike process; however, decisions made during our data collection may introduce limitations to our study. The online discussion posts in our analysis are all with more than 10 upvotes/downvotes during the time we collected the data, indicating a high level of activity. This approach, however, may exclude posts with relatively low activity but distinct viewpoints that differ from the majority. Additionally, all collected posts were retrieved from Meta Stack Exchange. Although Meta Stack Exchange serves as the primary site for moderators and users to discuss policy and regulations across all sites, individual sites may also provide unique insights specific to their communities. For example, opinions on Stack Overflow may differ from those on a humanities-focused site, such as English Language \& Usage. Future work may further explore any differences in viewpoints across sites. 

Of our interview participants, 13 out of 14 are current or former active users, moderators, or staff with at least 5 active years on Stack Exchange. We may miss perspectives from less active or newer community members by focusing on highly active users.

\section{Conclusion}

We examined why and how the SE strike took place and found that the long standing tensions between the community and SE, Inc., coupled with the sudden policy changes and restrictions on data dumps, led to the strike. As community members struggled to enact ``effective voice'' to bind the platform to support community needs, some ultimately resorted to exit, i.e. leaving the platform. In the aftermath, participants pursued different paths. Some resorted to alternative platforms, and some sought to preserve their knowledge. Building on our findings, we make recommendations for online communities, platform operators, and designers and developers to better navigate such tensions. We also discuss implications for the mechanism of community-driven data governance. We argue that the strike reflects community members’ desire for participatory governance not only over community rules and norms but also over the platform itself, expanding prior governance beyond “operational” and “collective” rules to include “constitutional” rules that define how platforms change.

\section{Acknowledgment}
The authors would like to thank the anonymous CSCW reviewers for their helpful feedback and
comments. The authors are also grateful to Kaylea Champion and Eric Fassbender for reading the draft of this work and providing thoughtful and constructive suggestions, and to Professor Ken Fleischmann for his early guidance on the interview protocol. The first author acknowledges support from the Bullard Fellowship, which funded this research, and appreciates Wendee Hsu for her time and suggestions on data collection. Most of all, we owe special gratitude to numerous moderators and knowledge contributors who voluntarily sustain the Stack Exchange communities, as well as to the 14 interview participants. Their dedication and insights profoundly shaped this study.




\bibliographystyle{ACM-Reference-Format}
\bibliography{zotero.bib, zotero_bak}

@article{frey_effective_2023,
	title = {Effective voice: {Beyond} exit and affect in online communities},
	volume = {25},
	issn = {1461-4448},
	shorttitle = {Effective voice},
	url = {https://doi.org/10.1177/14614448211044025},
	doi = {10.1177/14614448211044025},
	language = {EN},
	number = {9},
	urldate = {2025-11-26},
	journal = {New Media \& Society},
	author = {Frey, Seth and Schneider, Nathan},
	month = sep,
	year = {2023},
	note = {Publisher: SAGE Publications},
	pages = {2381--2398},
}

@article{frey_effective_2021,
	title = {Effective {Voice}: {Beyond} {Exit} and {Affect} in {Online} {Communities}},
	shorttitle = {Effective {Voice}},
	url = {http://arxiv.org/abs/2009.12470},
	doi = {10.48550/arXiv.2009.12470},
	urldate = {2025-09-16},
	author = {Frey, Seth and Schneider, Nathan},
	month = feb,
	year = {2021},
	note = {tex.ids= frey\_effective\_2023
arXiv: 2009.12470 [cs]
publisher: SAGE Publications},
}

@article{matias_preventing_2019,
	title = {Preventing harassment and increasing group participation through social norms in 2,190 online science discussions},
	volume = {116},
	url = {https://www.pnas.org/doi/full/10.1073/pnas.1813486116},
	doi = {10.1073/pnas.1813486116},
	number = {20},
	urldate = {2025-11-26},
	journal = {Proceedings of the National Academy of Sciences},
	author = {Matias, J. Nathan},
	month = may,
	year = {2019},
	note = {Publisher: Proceedings of the National Academy of Sciences},
	pages = {9785--9789},
}

@book{elster_ulysses_2000,
	address = {New York},
	title = {Ulysses {Unbound}: {Studies} in {Rationality}, {Precommitment}, and {Constraints}},
	shorttitle = {Ulysses {Unbound}},
	publisher = {Cambridge University Press},
	author = {Elster, Jon},
	year = {2000},
}

@inproceedings{vincent_data_2021,
	address = {Virtual Event Canada},
	title = {Data {Leverage}: {A} {Framework} for {Empowering} the {Public} in its {Relationship} with {Technology} {Companies}},
	isbn = {978-1-4503-8309-7},
	shorttitle = {Data {Leverage}},
	url = {https://dl.acm.org/doi/10.1145/3442188.3445885},
	doi = {10.1145/3442188.3445885},
	language = {en},
	urldate = {2025-11-16},
	booktitle = {Proceedings of the 2021 {ACM} {Conference} on {Fairness}, {Accountability}, and {Transparency}},
	publisher = {ACM},
	author = {Vincent, Nicholas and Li, Hanlin and Tilly, Nicole and Chancellor, Stevie and Hecht, Brent},
	month = mar,
	year = {2021},
	pages = {215--227},
}

@inproceedings{li_dimensions_2023,
	address = {Chicago IL USA},
	title = {The {Dimensions} of {Data} {Labor}: {A} {Road} {Map} for {Researchers}, {Activists}, and {Policymakers} to {Empower} {Data} {Producers}},
	isbn = {979-8-4007-0192-4},
	shorttitle = {The {Dimensions} of {Data} {Labor}},
	url = {https://dl.acm.org/doi/10.1145/3593013.3594070},
	doi = {10.1145/3593013.3594070},
	language = {en},
	urldate = {2025-11-16},
	booktitle = {2023 {ACM} {Conference} on {Fairness} {Accountability} and {Transparency}},
	publisher = {ACM},
	author = {Li, Hanlin and Vincent, Nicholas and Chancellor, Stevie and Hecht, Brent},
	month = jun,
	year = {2023},
	pages = {1151--1161},
}

@incollection{mannan_cooperative_2024,
	title = {Cooperative online communities},
	booktitle = {The {Routledge} {Handbook} of {Cooperative} {Economics} and {Management}},
	publisher = {Routledge},
	author = {Mannan, Morshed and Schneider, Nathan and Merk, Tara},
	year = {2024},
	note = {Num Pages: 9},
}

@article{lloyd_there_2025,
	title = {'{There} {Has} {To} {Be} a {Lot} {That} {We}'re {Missing}': {Moderating} {AI}-{Generated} {Content} on {Reddit}},
	volume = {9},
	issn = {2573-0142},
	shorttitle = {'{There} {Has} {To} {Be} a {Lot} {That} {We}'re {Missing}'},
	url = {https://dl.acm.org/doi/10.1145/3757445},
	doi = {10.1145/3757445},
	language = {en},
	number = {7},
	urldate = {2025-11-14},
	journal = {Proceedings of the ACM on Human-Computer Interaction},
	author = {Lloyd, Travis and Reagle, Joseph and Naaman, Mor},
	month = oct,
	year = {2025},
	pages = {1--24},
}

@misc{doctorow_how_2022,
	title = {How monopoly enshittified amazon},
	shorttitle = {Pluralistic},
	url = {https://pluralistic.net/2022/11/28/enshittification/},
	language = {american},
	urldate = {2025-09-15},
	journal = {Pluralistic: Daily links from Cory Doctorow},
	author = {Doctorow, Cory},
	month = aug,
	year = {2022},
}

@article{braun_using_2006,
	title = {Using thematic analysis in psychology},
	volume = {3},
	issn = {1478-0887},
	url = {https://doi.org/10.1191/1478088706qp063oa},
	doi = {10.1191/1478088706qp063oa},
	number = {2},
	urldate = {2025-09-17},
	journal = {Qualitative Research in Psychology},
	author = {Braun, Virginia and Clarke, Victoria},
	month = jan,
	year = {2006},
	note = {Publisher: Routledge
\_eprint: https://doi.org/10.1191/1478088706qp063oa},
	pages = {77--101},
}

@book{creamer_introduction_2018,
	title = {An {Introduction} to {Fully} {Integrated} {Mixed} {Methods} {Research}},
	isbn = {978-1-0718-0282-3},
	url = {https://methods.sagepub.com/book/mono/an-introduction-to-fully-integrated-mixed-methods-research/toc},
	language = {en},
	urldate = {2025-09-17},
	publisher = {SAGE Publications, Inc.},
	author = {Creamer, Elizabeth G.},
	year = {2018},
	doi = {10.4135/9781071802823},
}

@book{charmaz_constructing_2006,
	title = {Constructing {Grounded} {Theory}: {A} {Practical} {Guide} through {Qualitative} {Analysis}},
	isbn = {978-1-4462-0040-7},
	shorttitle = {Constructing {Grounded} {Theory}},
	language = {en},
	publisher = {SAGE},
	author = {Charmaz, Kathy},
	month = jan,
	year = {2006},
	note = {Google-Books-ID: 2ThdBAAAQBAJ},
}

@inproceedings{bryant_becoming_2005,
	address = {New York, NY, USA},
	series = {{GROUP} '05},
	title = {Becoming {Wikipedian}: transformation of participation in a collaborative online encyclopedia},
	isbn = {978-1-59593-223-5},
	shorttitle = {Becoming {Wikipedian}},
	url = {https://dl.acm.org/doi/10.1145/1099203.1099205},
	doi = {10.1145/1099203.1099205},
	urldate = {2025-09-16},
	booktitle = {Proceedings of the 2005 {ACM} {International} {Conference} on {Supporting} {Group} {Work}},
	publisher = {Association for Computing Machinery},
	author = {Bryant, Susan L. and Forte, Andrea and Bruckman, Amy},
	month = nov,
	year = {2005},
	pages = {1--10},
}

@inproceedings{jergensen_onion_2011,
	address = {New York, NY, USA},
	series = {{ESEC}/{FSE} '11},
	title = {The onion patch: migration in open source ecosystems},
	isbn = {978-1-4503-0443-6},
	shorttitle = {The onion patch},
	url = {https://dl.acm.org/doi/10.1145/2025113.2025127},
	doi = {10.1145/2025113.2025127},
	urldate = {2025-09-16},
	booktitle = {Proceedings of the 19th {ACM} {SIGSOFT} symposium and the 13th {European} conference on {Foundations} of software engineering},
	publisher = {Association for Computing Machinery},
	author = {Jergensen, Corey and Sarma, Anita and Wagstrom, Patrick},
	month = sep,
	year = {2011},
	pages = {70--80},
}

@misc{lave_situated_1991,
	title = {Situated {Learning}: {Legitimate} {Peripheral} {Participation}},
	shorttitle = {Situated {Learning}},
	url = {https://www.cambridge.org/highereducation/books/situated-learning/6915ABD21C8E4619F750A4D4ACA616CD},
	language = {en},
	urldate = {2025-09-16},
	journal = {Cambridge Aspire website},
	author = {Lave, Jean and Wenger, Etienne},
	month = sep,
	year = {1991},
	doi = {10.1017/CBO9780511815355},
	note = {ISBN: 9780511815355
Publisher: Cambridge University Press},
}

@inproceedings{zannettou_analyzing_2024,
	address = {New York, NY, USA},
	series = {{CHI} '24},
	title = {Analyzing {User} {Engagement} with {TikTok}'s {Short} {Format} {Video} {Recommendations} using {Data} {Donations}},
	isbn = {9798400703300},
	url = {https://dl.acm.org/doi/10.1145/3613904.3642433},
	doi = {10.1145/3613904.3642433},
	urldate = {2025-09-16},
	booktitle = {Proceedings of the 2024 {CHI} {Conference} on {Human} {Factors} in {Computing} {Systems}},
	publisher = {Association for Computing Machinery},
	author = {Zannettou, Savvas and Nemes-Nemeth, Olivia and Ayalon, Oshrat and Goetzen, Angelica and Gummadi, Krishna P. and Redmiles, Elissa M. and Roesner, Franziska},
	month = may,
	year = {2024},
	pages = {1--16},
}

@inproceedings{waltenberger_digital_2025,
	address = {New York, NY, USA},
	series = {Websci '25},
	title = {Digital {Sabotage} and {Inter}-{Community} {Dynamics}: {An} {Empirical} {Examination} of {Reddit}'s 2023 {Moderator} {Strike}},
	isbn = {9798400714832},
	shorttitle = {Digital {Sabotage} and {Inter}-{Community} {Dynamics}},
	url = {https://doi.org/10.1145/3717867.3717873},
	doi = {10.1145/3717867.3717873},
	urldate = {2025-09-14},
	booktitle = {Proceedings of the 17th {ACM} {Web} {Science} {Conference} 2025},
	publisher = {Association for Computing Machinery},
	author = {Waltenberger, Franz Xaver and Voggenreiter, Angelina and Pfeffer, Juergen},
	month = may,
	year = {2025},
	pages = {283--293},
}

@inproceedings{kleppmann_bluesky_2024,
	title = {Bluesky and the {AT} {Protocol}: {Usable} {Decentralized} {Social} {Media}},
	shorttitle = {Bluesky and the {AT} {Protocol}},
	url = {http://arxiv.org/abs/2402.03239},
	doi = {10.1145/3694809.3700740},
	urldate = {2025-08-29},
	booktitle = {Proceedings of the {ACM} {Conext}-2024 {Workshop} on the {Decentralization} of the {Internet}},
	author = {Kleppmann, Martin and Frazee, Paul and Gold, Jake and Graber, Jay and Holmgren, Daniel and Ivy, Devin and Johnson, Jeromy and Newbold, Bryan and Volpert, Jaz},
	month = dec,
	year = {2024},
	note = {arXiv:2402.03239 [cs]},
	pages = {1--7},
}

@inproceedings{gupta_being_2025,
	address = {Amsterdam Netherlands},
	title = {"{Being} a nanny isn't just caregiving": {An} {Analysis} of {How} {Nannies} {Seek} {Support} in {Online} {Communities} like /r/{Nanny}},
	isbn = {9798400713842},
	shorttitle = {"{Being} a nanny isn't just caregiving"},
	url = {https://dl.acm.org/doi/10.1145/3729176.3729181},
	doi = {10.1145/3729176.3729181},
	language = {en},
	urldate = {2025-08-23},
	booktitle = {Proceedings of the 4th {Annual} {Symposium} on {Human}-{Computer} {Interaction} for {Work}},
	publisher = {ACM},
	author = {Gupta, Meghna and Bhattacharya, Arpita and Kientz, Julie},
	month = jun,
	year = {2025},
	pages = {1--15},
}

@book{tarrow_power_1998,
	address = {Cambridge},
	edition = {2},
	series = {Cambridge {Studies} in {Comparative} {Politics}},
	title = {Power in {Movement}: {Social} {Movements} and {Contentious} {Politics}},
	shorttitle = {Power in {Movement}},
	url = {https://www.cambridge.org/core/books/power-in-movement/E9FC85E59075F0705549710D6A8BD858},
	urldate = {2025-05-14},
	publisher = {Cambridge University Press},
	author = {Tarrow, Sidney},
	year = {1998},
	doi = {10.1017/CBO9780511813245},
}

@misc{longpre_consent_2024,
	title = {Consent in {Crisis}: {The} {Rapid} {Decline} of the {AI} {Data} {Commons}},
	shorttitle = {Consent in {Crisis}},
	url = {http://arxiv.org/abs/2407.14933},
	doi = {10.48550/arXiv.2407.14933},
	urldate = {2025-05-13},
	publisher = {arXiv},
	author = {Longpre, Shayne and Mahari, Robert and Lee, Ariel and Lund, Campbell and Oderinwale, Hamidah and Brannon, William and Saxena, Nayan and Obeng-Marnu, Naana and South, Tobin and Hunter, Cole and Klyman, Kevin and Klamm, Christopher and Schoelkopf, Hailey and Singh, Nikhil and Cherep, Manuel and Anis, Ahmad and Dinh, An and Chitongo, Caroline and Yin, Da and Sileo, Damien and Mataciunas, Deividas and Misra, Diganta and Alghamdi, Emad and Shippole, Enrico and Zhang, Jianguo and Materzynska, Joanna and Qian, Kun and Tiwary, Kush and Miranda, Lester and Dey, Manan and Liang, Minnie and Hamdy, Mohammed and Muennighoff, Niklas and Ye, Seonghyeon and Kim, Seungone and Mohanty, Shrestha and Gupta, Vipul and Sharma, Vivek and Chien, Vu Minh and Zhou, Xuhui and Li, Yizhi and Xiong, Caiming and Villa, Luis and Biderman, Stella and Li, Hanlin and Ippolito, Daphne and Hooker, Sara and Kabbara, Jad and Pentland, Sandy},
	month = jul,
	year = {2024},
	note = {arXiv:2407.14933 [cs]},
}

@misc{reeves_exploring_2024,
	title = {Exploring the {Impact} of {ChatGPT} on {Wikipedia} {Engagement}},
	url = {http://arxiv.org/abs/2405.10205},
	doi = {10.48550/arXiv.2405.10205},
	urldate = {2025-05-13},
	publisher = {arXiv},
	author = {Reeves, Neal and Yin, Wenjie and Simperl, Elena},
	month = may,
	year = {2024},
	note = {arXiv:2405.10205 [cs]},
}

@article{li_impacts_2024,
	title = {Impacts of generative {AI} on user contributions: evidence from a coding {Q} \&{A} platform},
	copyright = {2024 The Author(s)},
	issn = {1573-059X},
	shorttitle = {Impacts of generative {AI} on user contributions},
	url = {https://link.springer.com/article/10.1007/s11002-024-09747-1},
	doi = {10.1007/s11002-024-09747-1},
	language = {en},
	urldate = {2025-05-13},
	journal = {Marketing Letters},
	author = {Li, Xinyu and Kim, Keongtae},
	month = sep,
	year = {2024},
	note = {Company: Springer
Distributor: Springer
Institution: Springer
Label: Springer
Publisher: Springer US},
	pages = {1--15},
}

@article{vetter_endangered_2025,
	title = {An endangered species: how {LLMs} threaten {Wikipedia}’s sustainability},
	copyright = {2025 The Author(s)},
	issn = {1435-5655},
	shorttitle = {An endangered species},
	url = {https://link.springer.com/article/10.1007/s00146-025-02199-9},
	doi = {10.1007/s00146-025-02199-9},
	language = {en},
	urldate = {2025-05-13},
	journal = {AI \& SOCIETY},
	author = {Vetter, Matthew A. and Jiang, Jialei and McDowell, Zachary J.},
	month = feb,
	year = {2025},
	note = {Company: Springer
Distributor: Springer
Institution: Springer
Label: Springer
Publisher: Springer London},
	pages = {1--14},
}

@article{wagner_death_2025,
	title = {Death by {AI}: {Will} large language models diminish {Wikipedia}?},
	volume = {76},
	copyright = {© 2025 The Author(s). Journal of the Association for Information Science and Technology published by Wiley Periodicals LLC on behalf of Association for Information Science and Technology.},
	issn = {2330-1643},
	shorttitle = {Death by {AI}},
	url = {https://onlinelibrary.wiley.com/doi/abs/10.1002/asi.24975},
	doi = {10.1002/asi.24975},
	language = {en},
	number = {5},
	urldate = {2025-05-13},
	journal = {Journal of the Association for Information Science and Technology},
	author = {Wagner, Christian and Jiang, Ling},
	year = {2025},
	note = {\_eprint: https://onlinelibrary.wiley.com/doi/pdf/10.1002/asi.24975},
	pages = {743--751},
}

@inproceedings{brooks_rise_2024,
	address = {Miami, Florida, USA},
	title = {The {Rise} of {AI}-{Generated} {Content} in {Wikipedia}},
	url = {https://aclanthology.org/2024.wikinlp-1.12},
	doi = {10.18653/v1/2024.wikinlp-1.12},
	language = {en},
	urldate = {2025-05-13},
	booktitle = {Proceedings of the {First} {Workshop} on {Advancing} {Natural} {Language} {Processing} for {Wikipedia}},
	publisher = {Association for Computational Linguistics},
	author = {Brooks, Creston and Eggert, Samuel and Peskoff, Denis},
	year = {2024},
	pages = {67--79},
}

@misc{lyu_wikipedia_2025,
	title = {Wikipedia {Contributions} in the {Wake} of {ChatGPT}},
	url = {http://arxiv.org/abs/2503.00757},
	doi = {10.48550/arXiv.2503.00757},
	urldate = {2025-05-13},
	publisher = {arXiv},
	author = {Lyu, Liang and Siderius, James and Li, Hannah and Acemoglu, Daron and Huttenlocher, Daniel and Ozdaglar, Asuman},
	month = mar,
	year = {2025},
	note = {arXiv:2503.00757 [cs]},
}

@article{shaw_laboratories_2014,
	title = {Laboratories of {Oligarchy}? {How} the {Iron} {Law} {Extends} to {Peer} {Production}},
	volume = {64},
	issn = {0021-9916},
	shorttitle = {Laboratories of {Oligarchy}?},
	url = {https://doi.org/10.1111/jcom.12082},
	doi = {10.1111/jcom.12082},
	number = {2},
	urldate = {2025-05-13},
	journal = {Journal of Communication},
	author = {Shaw, Aaron and Hill, Benjamin M.},
	month = apr,
	year = {2014},
	pages = {215--238},
}

@article{teblunthuis_effects_2021,
	title = {Effects of {Algorithmic} {Flagging} on {Fairness}: {Quasi}-experimental {Evidence} from {Wikipedia}},
	volume = {5},
	shorttitle = {Effects of {Algorithmic} {Flagging} on {Fairness}},
	url = {https://dl.acm.org/doi/10.1145/3449130},
	doi = {10.1145/3449130},
	number = {CSCW1},
	urldate = {2025-05-13},
	journal = {Proc. ACM Hum.-Comput. Interact.},
	author = {TeBlunthuis, Nathan and Hill, Benjamin Mako and Halfaker, Aaron},
	month = apr,
	year = {2021},
	pages = {56:1--56:27},
}

@book{schneider_governable_2024,
	address = {Oakland},
	title = {Governable spaces: democratic design for online life},
	isbn = {978-0-520-39395-0},
	shorttitle = {Governable spaces},
	url = {https://search.ebscohost.com/login.aspx?direct=true&scope=site&db=nlebk&db=nlabk&AN=3777996},
	language = {eng},
	urldate = {2025-05-12},
	publisher = {University of California Press},
	author = {Schneider, Nathan},
	collaborator = {Medić, Darija},
	year = {2024},
	note = {OCLC: 1397061567},
}

@article{grimmelmann_virtues_2015,
	title = {The {Virtues} of {Moderation}},
	volume = {17},
	url = {https://heinonline.org/HOL/P?h=hein.journals/yjolt17&i=42},
	language = {eng},
	urldate = {2025-05-12},
	journal = {Yale Journal of Law and Technology},
	author = {Grimmelmann, James},
	year = {2015},
	pages = {42--109},
}

@article{seering_metaphors_2022,
	title = {Metaphors in moderation},
	volume = {24},
	issn = {1461-4448},
	url = {https://doi.org/10.1177/1461444820964968},
	doi = {10.1177/1461444820964968},
	language = {EN},
	number = {3},
	urldate = {2025-05-12},
	journal = {New Media \& Society},
	author = {Seering, Joseph and Kaufman, Geoff and Chancellor, Stevie},
	month = mar,
	year = {2022},
	note = {Publisher: SAGE Publications},
	pages = {621--640},
}

@article{burtch_consequences_2024,
	title = {The consequences of generative {AI} for online knowledge communities},
	volume = {14},
	copyright = {2024 The Author(s)},
	issn = {2045-2322},
	url = {https://www.nature.com/articles/s41598-024-61221-0},
	doi = {10.1038/s41598-024-61221-0},
	language = {en},
	number = {1},
	urldate = {2025-05-12},
	journal = {Scientific Reports},
	author = {Burtch, Gordon and Lee, Dokyun and Chen, Zhichen},
	month = may,
	year = {2024},
	note = {Publisher: Nature Publishing Group},
	pages = {10413},
}

@article{jhaver_decentralizing_2023,
	title = {Decentralizing {Platform} {Power}: {A} {Design} {Space} of {Multi}-{Level} {Governance} in {Online} {Social} {Platforms}},
	volume = {9},
	issn = {2056-3051},
	shorttitle = {Decentralizing {Platform} {Power}},
	url = {https://doi.org/10.1177/20563051231207857},
	doi = {10.1177/20563051231207857},
	language = {EN},
	number = {4},
	urldate = {2025-05-11},
	journal = {Social Media + Society},
	author = {Jhaver, Shagun and Frey, Seth and Zhang, Amy X.},
	month = oct,
	year = {2023},
	note = {Publisher: SAGE Publications Ltd},
	pages = {20563051231207857},
}

@article{zuckerman_community_2023,
	title = {From {Community} {Governance} to {Customer} {Service} and {Back} {Again}: {Re}-{Examining} {Pre}-{Web} {Models} of {Online} {Governance} to {Address} {Platforms}’ {Crisis} of {Legitimacy}},
	volume = {9},
	issn = {2056-3051},
	shorttitle = {From {Community} {Governance} to {Customer} {Service} and {Back} {Again}},
	url = {https://doi.org/10.1177/20563051231196864},
	doi = {10.1177/20563051231196864},
	language = {EN},
	number = {3},
	urldate = {2025-05-11},
	journal = {Social Media + Society},
	author = {Zuckerman, Ethan and Rajendra-Nicolucci, Chand},
	month = jul,
	year = {2023},
	note = {Publisher: SAGE Publications Ltd},
	pages = {20563051231196864},
}

@inproceedings{kuo_policycraft_2025,
	address = {New York, NY, USA},
	series = {{CHI} '25},
	title = {{PolicyCraft}: {Supporting} {Collaborative} and {Participatory} {Policy} {Design} through {Case}-{Grounded} {Deliberation}},
	isbn = {9798400713941},
	shorttitle = {{PolicyCraft}},
	url = {https://dl.acm.org/doi/10.1145/3706598.3713865},
	doi = {10.1145/3706598.3713865},
	urldate = {2025-05-11},
	booktitle = {Proceedings of the 2025 {CHI} {Conference} on {Human} {Factors} in {Computing} {Systems}},
	publisher = {Association for Computing Machinery},
	author = {Kuo, Tzu-Sheng and Chen, Quan Ze and Zhang, Amy X. and Hsieh, Jane and Zhu, Haiyi and Holstein, Kenneth},
	month = apr,
	year = {2025},
	pages = {1--24},
}

@inproceedings{zhang_policykit_2020,
	address = {New York, NY, USA},
	series = {{UIST} '20},
	title = {{PolicyKit}: {Building} {Governance} in {Online} {Communities}},
	isbn = {978-1-4503-7514-6},
	shorttitle = {{PolicyKit}},
	url = {https://doi.org/10.1145/3379337.3415858},
	doi = {10.1145/3379337.3415858},
	urldate = {2025-05-11},
	booktitle = {Proceedings of the 33rd {Annual} {ACM} {Symposium} on {User} {Interface} {Software} and {Technology}},
	publisher = {Association for Computing Machinery},
	author = {Zhang, Amy X. and Hugh, Grant and Bernstein, Michael S.},
	month = oct,
	year = {2020},
	pages = {365--378},
}

@article{weld_making_2024,
	title = {Making {Online} {Communities} ‘{Better}’: {A} {Taxonomy} of {Community} {Values} on {Reddit}},
	volume = {18},
	issn = {2334-0770, 2162-3449},
	shorttitle = {Making {Online} {Communities} ‘{Better}’},
	url = {https://ojs.aaai.org/index.php/ICWSM/article/view/31413},
	doi = {10.1609/icwsm.v18i1.31413},
	language = {en},
	urldate = {2025-05-11},
	journal = {Proceedings of the International AAAI Conference on Web and Social Media},
	author = {Weld, Galen and Zhang, Amy X. and Althoff, Tim},
	month = may,
	year = {2024},
	pages = {1611--1633},
}

@article{matias_civic_2019,
	title = {The civic labor of volunteer moderators online},
	volume = {5},
	issn = {2056-3051, 2056-3051},
	url = {http://journals.sagepub.com/doi/10.1177/2056305119836778},
	doi = {https://doi.org/10.1177/2056305119836778},
	language = {en},
	number = {2},
	urldate = {2019-07-23},
	journal = {Social Media + Society},
	author = {Matias, J. Nathan},
	month = apr,
	year = {2019},
	note = {tex.ids= matias\_civic\_2019-1},
	pages = {1--12},
}

@misc{mithical_moderation_2023,
	type = {Forum post},
	title = {Moderation {Strike}: {Stack} {Overflow}, {Inc}. cannot consistently ignore, mistreat, and malign its volunteers},
	shorttitle = {Moderation {Strike}},
	url = {https://meta.stackexchange.com/q/389811},
	urldate = {2025-05-11},
	journal = {Meta Stack Exchange},
	author = {Mithical},
	month = jun,
	year = {2023},
}

@misc{mithical_answer_2020,
	title = {Answer to "{Can} a site enforce its own rules above and beyond the {Code} of {Conduct}?"},
	shorttitle = {Answer to "{Can} a site enforce its own rules above and beyond the {Code} of {Conduct}?},
	url = {https://meta.stackexchange.com/a/341599/1564075},
	urldate = {2025-05-08},
	journal = {Meta Stack Exchange},
	author = {Mithical},
	month = jan,
	year = {2020},
}

@inproceedings{spaa_understanding_2019,
	address = {Glasgow Scotland Uk},
	title = {Understanding the {Boundaries} between {Policymaking} and {HCI}},
	isbn = {978-1-4503-5970-2},
	url = {https://dl.acm.org/doi/10.1145/3290605.3300314},
	doi = {10.1145/3290605.3300314},
	language = {en},
	urldate = {2025-05-01},
	booktitle = {Proceedings of the 2019 {CHI} {Conference} on {Human} {Factors} in {Computing} {Systems}},
	publisher = {ACM},
	author = {Spaa, Anne and Durrant, Abigail and Elsden, Chris and Vines, John},
	month = may,
	year = {2019},
	pages = {1--15},
}

@inproceedings{matias_civilservant_2018,
	address = {New York, NY, USA},
	series = {{CHI} '18},
	title = {{CivilServant}: {Community}-{Led} {Experiments} in {Platform} {Governance}},
	isbn = {978-1-4503-5620-6},
	shorttitle = {{CivilServant}},
	url = {https://doi.org/10.1145/3173574.3173583},
	doi = {10.1145/3173574.3173583},
	urldate = {2025-04-30},
	booktitle = {Proceedings of the 2018 {CHI} {Conference} on {Human} {Factors} in {Computing} {Systems}},
	publisher = {Association for Computing Machinery},
	author = {Matias, J. Nathan and Mou, Merry},
	month = apr,
	year = {2018},
	pages = {1--13},
}

@book{hirschman_exit_1970,
	title = {Exit, {Voice}, and {Loyalty}: {Responses} to {Decline} in {Firms}, {Organizations}, and {States}},
	isbn = {978-0-674-27660-4},
	shorttitle = {Exit, {Voice}, and {Loyalty}},
	language = {en},
	publisher = {Harvard University Press},
	author = {Hirschman, Albert O.},
	year = {1970},
	note = {Google-Books-ID: vYO6sDvjvcgC},
}

@article{newell_user_2016,
	title = {User {Migration} in {Online} {Social} {Networks}: {A} {Case} {Study} on {Reddit} {During} a {Period} of {Community} {Unrest}},
	volume = {10},
	copyright = {Copyright (c) 2021 Proceedings of the International AAAI Conference on Web and Social Media},
	issn = {2334-0770},
	shorttitle = {User {Migration} in {Online} {Social} {Networks}},
	url = {https://ojs.aaai.org/index.php/ICWSM/article/view/14750},
	doi = {10.1609/icwsm.v10i1.14750},
	language = {en},
	number = {1},
	urldate = {2025-03-20},
	journal = {Proceedings of the International AAAI Conference on Web and Social Media},
	author = {Newell, Edward and Jurgens, David and Saleem, Haji and Vala, Hardik and Sassine, Jad and Armstrong, Caitrin and Ruths, Derek},
	year = {2016},
	note = {Number: 1},
	pages = {279--288},
}

@inproceedings{centivany_popcorn_2016,
	address = {New York, NY, USA},
	series = {{CHI} '16},
	title = {"{Popcorn} {Tastes} {Good}": {Participatory} {Policymaking} and {Reddit}'s},
	isbn = {978-1-4503-3362-7},
	shorttitle = {"{Popcorn} {Tastes} {Good}"},
	url = {https://dl.acm.org/doi/10.1145/2858036.2858516},
	doi = {10.1145/2858036.2858516},
	urldate = {2025-04-28},
	booktitle = {Proceedings of the 2016 {CHI} {Conference} on {Human} {Factors} in {Computing} {Systems}},
	publisher = {Association for Computing Machinery},
	author = {Centivany, Alissa and Glushko, Bobby},
	month = may,
	year = {2016},
	pages = {1126--1137},
}

@book{gowder_networked_2023,
	edition = {1},
	title = {The {Networked} {Leviathan}: {For} {Democratic} {Platforms}},
	copyright = {https://www.cambridge.org/core/terms},
	isbn = {978-1-108-97543-8 978-1-108-83862-7 978-1-108-97190-4},
	shorttitle = {The {Networked} {Leviathan}},
	url = {https://www.cambridge.org/core/product/identifier/9781108975438/type/book},
	language = {en},
	urldate = {2025-04-27},
	publisher = {Cambridge University Press},
	author = {Gowder, Paul},
	month = aug,
	year = {2023},
	doi = {10.1017/9781108975438},
}

@article{schneider_admins_2022,
	title = {Admins, mods, and benevolent dictators for life: {The} implicit feudalism of online communities},
	volume = {24},
	issn = {1461-4448},
	shorttitle = {Admins, mods, and benevolent dictators for life},
	url = {https://doi.org/10.1177/1461444820986553},
	doi = {10.1177/1461444820986553},
	language = {EN},
	number = {9},
	urldate = {2025-04-26},
	journal = {New Media \& Society},
	author = {Schneider, Nathan},
	month = sep,
	year = {2022},
	note = {Publisher: SAGE Publications},
	pages = {1965--1985},
}

@article{frey_this_2019,
	title = {"{This} {Place} {Does} {What} {It} {Was} {Built} {For}": {Designing} {Digital} {Institutions} for {Participatory} {Change}},
	volume = {3},
	shorttitle = {"{This} {Place} {Does} {What} {It} {Was} {Built} {For}"},
	url = {https://dl.acm.org/doi/10.1145/3359134},
	doi = {10.1145/3359134},
	number = {CSCW},
	urldate = {2025-04-26},
	journal = {Proc. ACM Hum.-Comput. Interact.},
	author = {Frey, Seth and Krafft, P. M. and Keegan, Brian C.},
	month = nov,
	year = {2019},
	pages = {32:1--32:31},
}

@article{jemielniak_wikimedia_2016,
	title = {Wikimedia movement governance: the limits of a-hierarchical organization},
	volume = {29},
	copyright = {https://www.emerald.com/insight/site-policies},
	issn = {0953-4814},
	shorttitle = {Wikimedia movement governance},
	url = {https://www.emerald.com/insight/content/doi/10.1108/JOCM-07-2013-0138/full/html},
	doi = {10.1108/JOCM-07-2013-0138},
	language = {en},
	number = {3},
	urldate = {2025-04-23},
	journal = {Journal of Organizational Change Management},
	author = {Jemielniak, Dariusz},
	month = may,
	year = {2016},
	note = {tex.ids= jemielniak\_wikimedia\_2016-1},
	pages = {361--378},
}

@inproceedings{cavusoglu_can_2015,
	address = {New York, NY, USA},
	series = {{CSCW}'15 {Companion}},
	title = {Can {Gamification} {Motivate} {Voluntary} {Contributions}? {The} {Case} of {StackOverflow} {Q}\&{A} {Community}},
	isbn = {978-1-4503-2946-0},
	shorttitle = {Can {Gamification} {Motivate} {Voluntary} {Contributions}?},
	url = {https://doi.org/10.1145/2685553.2698999},
	doi = {10.1145/2685553.2698999},
	urldate = {2025-04-08},
	booktitle = {Proceedings of the 18th {ACM} {Conference} {Companion} on {Computer} {Supported} {Cooperative} {Work} \& {Social} {Computing}},
	publisher = {Association for Computing Machinery},
	author = {Cavusoglu, Huseyin and Li, Zhuolun and Huang, Ke-Wei},
	month = feb,
	year = {2015},
	pages = {171--174},
}

@article{fiesler_moving_2020,
	title = {Moving {Across} {Lands}: {Online} {Platform} {Migration} in {Fandom} {Communities}},
	volume = {4},
	issn = {2573-0142},
	shorttitle = {Moving {Across} {Lands}},
	url = {https://dl.acm.org/doi/10.1145/3392847},
	doi = {10.1145/3392847},
	language = {en},
	number = {CSCW1},
	urldate = {2025-02-26},
	journal = {Proceedings of the ACM on Human-Computer Interaction},
	author = {Fiesler, Casey and Dym, Brianna},
	month = may,
	year = {2020},
	pages = {1--25},
}

@inproceedings{matias_going_2016,
	address = {San Jose California USA},
	title = {Going {Dark}: {Social} {Factors} in {Collective} {Action} {Against} {Platform} {Operators} in the {Reddit} {Blackout}},
	isbn = {978-1-4503-3362-7},
	shorttitle = {Going {Dark}},
	url = {https://dl.acm.org/doi/10.1145/2858036.2858391},
	doi = {10.1145/2858036.2858391},
	language = {en},
	urldate = {2024-07-14},
	booktitle = {Proceedings of the 2016 {CHI} {Conference} on {Human} {Factors} in {Computing} {Systems}},
	publisher = {ACM},
	author = {Matias, J. Nathan},
	month = may,
	year = {2016},
	pages = {1138--1151},
}

@online{mathias_undeclared_2025,
	title = {Undeclared {AI}-generated text in Wikipedia: A tale of caution Wikimania 2025},
	url = {https://wikimedia.eventyay.com/talk/wikimania2025/talk/HZEQZW/},
	shorttitle = {Undeclared {AI}-generated text in Wikipedia},
	urldate = {2025-11-18},
	date = {2025-08-06},
	langid = {english},
}

@inproceedings{ajmaniSystematicReviewEthics2023, author = {Ajmani, Leah Hope and Chancellor, Stevie and Mehta, Bijal and Fiesler, Casey and Zimmer, Michael and De Choudhury, Munmun}, title = {A Systematic Review of Ethics Disclosures in Predictive Mental Health Research}, year = {2023}, isbn = {9798400701924}, publisher = {Association for Computing Machinery}, address = {New York, NY, USA}, url = {https://doi.org/10.1145/3593013.3594082}, doi = {10.1145/3593013.3594082}, booktitle = {Proceedings of the 2023 ACM Conference on Fairness, Accountability, and Transparency}, pages = {1311–1323}, numpages = {13}, location = {Chicago, IL, USA}, series = {FAccT '23} }

@inproceedings{ajmani2024data,
  title={Data Agency Theory: A Precise Theory of Justice for AI Applications},
  author={Ajmani, Leah and Stapleton, Logan and Houtti, Mo and Chancellor, Stevie},
  booktitle={The 2024 ACM Conference on Fairness, Accountability, and Transparency},
  pages={631--641},
  year={2024},
  url = {https://doi.org/10.1145/3630106.365893}
}

@ARTICLE{Gilbert2020-hu,
  title     = "``{I} run the world's largest historical outreach project and
               it's on a cesspool of a website.'' Moderating a Public
               Scholarship Site on Reddit: A Case Study of r/{AskHistorians}",
  author    = "Gilbert, Sarah A",
  journal   = "Proc. ACM Hum. Comput. Interact.",
  publisher = "Association for Computing Machinery (ACM)",
  address   = "New York, NY, USA",
  volume    =  4,
  number    = "CSCW1",
  pages     = "1--27",
  month     =  may,
  year      =  2020,
  url       =  {https://dl.acm.org/doi/10.1145/3392822},
  language  = "en"
}

@ARTICLE{Gilbert2023-pg,
  title     = "Towards intersectional moderation: An alternative model of
               moderation built on care and power",
  author    = "Gilbert, Sarah",
  journal   = "Proc. ACM Hum. Comput. Interact.",
  publisher = "Association for Computing Machinery (ACM)",
  address   = "New York, NY, USA",
  volume    =  7,
  number    = "CSCW2",
  pages     = "1--32",
  month     =  sep,
  year      =  2023,
  language  = "en"
}

@INPROCEEDINGS{Geiger2010-ny,
  title     = "The work of sustaining order in wikipedia: the banning of a
               vandal",
  author    = "Geiger, R Stuart and Ribes, David",
  booktitle = "Proceedings of the 2010 ACM conference on Computer supported
               cooperative work",
  publisher = "ACM",
  address   = "New York, NY, USA",
  pages     = "117--126",
  series    = "CSCW '10",
  month     =  feb,
  year      =  {2010},
  url       = {https://dl.acm.org/doi/10.1145/1718918.1718941}
}

@article{Vincent2021-xl, author = {Vincent, Nicholas and Hecht, Brent}, title = {A Deeper Investigation of the Importance of Wikipedia Links to Search Engine Results}, year = {2021}, issue_date = {April 2021}, publisher = {Association for Computing Machinery}, address = {New York, NY, USA}, volume = {5}, number = {CSCW1}, url = {https://doi.org/10.1145/3449078}, doi = {10.1145/3449078}, journal = {Proc. ACM Hum.-Comput. Interact.}, month = apr, articleno = {4}, numpages = {15} }

@online{license_ccbysa_24,
  author = {CREATIVE COMMONS},
  title = {ATTRIBUTION-SHAREALIKE 4.0 INTERNATIONAL DEED},
  url = {https://creativecommons.org/licenses/by-sa/4.0/},
  urldate = {n/a},
}

@online{so_offline_22,
  author = {Ben Popper},
  year = {2022},
  title = {Introducing the Overflow Offline project},
  url = {https://stackoverflow.blog/2022/10/20/introducing-the-overflow-offline-project/},
  urldate = {n/a},
}

@online{se_history_24,
  author = {Janhrach},
  title = {Stack Exchange},
  year = {2023},
  url = {https://en.wikipedia.org/wiki/Stack_Exchange},
  note = {Updated October 18, 2023}
}

@misc{li_all_2022,
      title={All That's Happening behind the Scenes: Putting the Spotlight on Volunteer Moderator Labor in Reddit}, 
      author={Hanlin Li and Brent Hecht and Stevie Chancellor},
      year={2022},
      eprint={2205.14529},
      archivePrefix={arXiv},
      primaryClass={cs.HC},
      url={https://arxiv.org/abs/2205.14529}, 
}

@article{liMeasuringMonetaryValue2022,
  title = {Measuring the {{Monetary Value}} of {{Online Volunteer Work}}},
  author = {Li, Hanlin and Hecht, Brent and Chancellor, Stevie},
  year = 2022,
  month = May,
  journaltitle = {Proceedings of the International AAAI Conference on Web and Social Media},
  volume = {16},
  pages = {596--606},
  issn = {2334-0770},
  doi = {10.1609/icwsm.v16i1.19318},
  url = {https://ojs.aaai.org/index.php/ICWSM/article/view/19318},
  urldate = {2024-07-22},
  langid = {english}
}

@article{mcinnisReportingCommunityBeat2021, author = {McInnis, Brian and Ajmani, Leah and Sun, Lu and Hou, Yiwen and Zeng, Ziwen and Dow, Steven P.}, title = {Reporting the Community Beat: Practices for Moderating Online Discussion at a News Website}, year = {2021}, issue_date = {October 2021}, publisher = {Association for Computing Machinery}, address = {New York, NY, USA}, volume = {5}, number = {CSCW2}, url = {https://doi.org/10.1145/3476074}, doi = {10.1145/3476074}, journal = {Proc. ACM Hum.-Comput. Interact.}, month = oct, articleno = {333}, numpages = {25} }

@book{saldanaCodingManualQualitative2013,
  title = {The Coding Manual for Qualitative Researchers},
  author = {Saldaña, Johnny},
  year = {2013},
  edition = {2nd ed},
  publisher = {SAGE},
  location = {Los Angeles},
  isbn = {978-1-4462-4736-5 978-1-4462-4737-2},
  langid = {english},
  pagetotal = {303}
}

@journal{longpre2024data,
  title={Data Authenticity, Consent, and Provenance for AI Are All Broken: What Will It Take to Fix Them?},
  author={Longpre, Shayne and Mahari, Robert and Obeng-Marnu, Naana and Brannon, William and South, Tobin and Kabbara, Jad and Pentland, Sandy},
  year={2024},
  publisher={MIT},
  url = {https://doi.org/10.48550/arXiv.2404.12691}
}

@article{longpre2023data,
  title={The data provenance initiative: A large scale audit of dataset licensing \& attribution in ai},
  author={Longpre, Shayne and Mahari, Robert and Chen, Anthony and Obeng-Marnu, Naana and Sileo, Damien and Brannon, William and Muennighoff, Niklas and Khazam, Nathan and Kabbara, Jad and Perisetla, Kartik and others},
  journal={arXiv preprint arXiv:2310.16787},
  year={2023},
  url = {url = {https://doi.org/10.1038/s42256-024-00878-8}}
}

@article{gebru2021datasheets,
  title={Datasheets for datasets},
  author={Gebru, Timnit and Morgenstern, Jamie and Vecchione, Briana and Vaughan, Jennifer Wortman and Wallach, Hanna and Iii, Hal Daum{\'e} and Crawford, Kate},
  journal={Communications of the ACM},
  volume={64},
  number={12},
  pages={86--92},
  year={2021},
  url = {https://dl.acm.org/doi/10.1145/3458723},
  publisher={ACM New York, NY, USA}
}

@inproceedings{wang_quality_2015,
author = {Warncke-Wang, Morten and Ayukaev, Vladislav R. and Hecht, Brent and Terveen, Loren G.},
title = {The Success and Failure of Quality Improvement Projects in Peer Production Communities},
year = {2015},
isbn = {9781450329224},
publisher = {Association for Computing Machinery},
address = {New York, NY, USA},
url = {https://doi.org/10.1145/2675133.2675241},
doi = {10.1145/2675133.2675241},
booktitle = {Proceedings of the 18th ACM Conference on Computer Supported Cooperative Work \& Social Computing},
pages = {743–756},
numpages = {14},
location = {Vancouver, BC, Canada},
series = {CSCW '15}
}

@article{dabbish_transparency_2014,
  author       = {Laura Dabbish and
                  H. Colleen Stuart and
                  Jason Tsay and
                  James D. Herbsleb},
  title        = {Transparency and Coordination in Peer Production},
  journal      = {CoRR},
  volume       = {abs/1407.0377},
  year         = {2014},
  url          = {http://arxiv.org/abs/1407.0377},
  eprinttype    = {arXiv},
  eprint       = {1407.0377},
  bibsource    = {dblp computer science bibliography, https://dblp.org}
}

@misc{kuo_policycraft_2024,
      title={PolicyCraft: Supporting Collaborative and Participatory Policy Design through Case-Grounded Deliberation}, 
      author={Tzu-Sheng Kuo and Quan Ze Chen and Amy X. Zhang and Jane Hsieh and Haiyi Zhu and Kenneth Holstein},
      year={2024},
      eprint={2409.15644},
      archivePrefix={arXiv},
      primaryClass={cs.HC},
      url={https://doi.org/10.48550/arXiv.2409.15644}, 
}

@inproceedings{zhangQualitativeAnalysisContent,
  title={Qualitative Analysis of Content by},
  author={Yan Zhang and Barbara M. Wildemuth},
  year={2005},
  url={https://api.semanticscholar.org/CorpusID:16523714}
}

@online{atwood_theory_2009,
  title = {A {{Theory}} of {{Moderation}} - {{Stack Overflow}}},
  date = {2009-05-18},
  author = {Atwood, Jeff},
  year = {2009},
  url = {https://stackoverflow.blog/2009/05/18/a-theory-of-moderation/},
  urldate = {2024-10-30},
note = {Retrieved October 29, 2024 from https://stackoverflow.blog/2009/05/18/a-theory-of-moderation/},
  langid = {english}
}

@article{sharma_consensus_2022, author = {Sharma, Nirwan and Colucci-Gray, Laura and van der Wal, Ren\'{e} and Siddharthan, Advaith}, title = {Consensus Building in On-Line Citizen Science}, year = {2022}, issue_date = {November 2022}, publisher = {Association for Computing Machinery}, address = {New York, NY, USA}, volume = {6}, number = {CSCW2}, url = {https://doi.org/10.1145/3555535}, doi = {10.1145/3555535}, journal = {Proc. ACM Hum.-Comput. Interact.}, month = nov, articleno = {434}, numpages = {26} }

@article{vincent_measuring_2019,
    title = {Measuring the {Importance} of {User}-{Generated} {Content} to {Search} {Engines}},
    volume = {13},
    copyright = {Copyright (c) 2019 Association for the Advancement of Artificial Intelligence},
    issn = {2334-0770},
    url = {https://ojs.aaai.org/index.php/ICWSM/article/view/3248},
    doi = {10.1609/icwsm.v13i01.3248},
    language = {en},
    urldate = {2024-07-29},
    journal = {Proceedings of the International AAAI Conference on Web and Social Media},
    author = {Vincent, Nicholas and Johnson, Isaac and Sheehan, Patrick and Hecht, Brent},
    month = jul,
    year = {2019},
    pages = {505--516},
}

@inproceedings{Vines_Configuration_2013,
author = {Vines, John and Clarke, Rachel and Wright, Peter and McCarthy, John and Olivier, Patrick},
title = {Configuring participation: on how we involve people in design},
year = {2013},
isbn = {9781450318990},
publisher = {Association for Computing Machinery},
address = {New York, NY, USA},
url = {https://doi.org/10.1145/2470654.2470716},
doi = {10.1145/2470654.2470716},
booktitle = {Proceedings of the SIGCHI Conference on Human Factors in Computing Systems},
pages = {429–438},
numpages = {10},
location = {Paris, France},
series = {CHI '13}
}

@misc{Wettig_2025_Organize,
      title={Organize the Web: Constructing Domains Enhances Pre-Training Data Curation}, 
      author={Alexander Wettig and Kyle Lo and Sewon Min and Hannaneh Hajishirzi and Danqi Chen and Luca Soldaini},
      year={2025},
      eprint={2502.10341},
      archivePrefix={arXiv},
      primaryClass={cs.CL},
      url={https://arxiv.org/abs/2502.10341}, 
}
\end{document}